%% file: paper.tex
\documentstyle{mn}
\input{psfig}
\input{macros_vdbosch}
\begin{document}


\title[Towards a Concordant Model of Halo Occupation Statistics]
      {Towards a Concordant Model of Halo Occupation Statistics}
\author[van den Bosch et al.]
       {\parbox[t]{\textwidth}{
        Frank C. van den Bosch$^{1}$\thanks{E-mail: vdbosch@mpia.de}, 
        Xiaohu Yang$^{2}$, H.J. Mo$^{3}$, Simone M. Weinmann$^{4}$,\\ 
        Andrea V. Macci\`o$^{4}$, Surhud More$^{1}$, Marcello Cacciato$^{1}$, 
        Ramin Skibba$^{1}$, Kang Xi$^{1}$}\\
        \vspace*{3pt} \\
       $^1$Max-Planck-Institute for Astronomy, K\"onigstuhl 17, D-69117
           Heidelberg, Germany\\
       $^2$Shanghai Astronomical Observatory; the Partner Group of MPA,
           Nandan Road 80,  Shanghai 200030, China\\
       $^3$Department of Astronomy, University of Massachusetts,
           Amherst MA 01003-9305, USA\\
       $^4$Institute for Theoretical Physics, University of Z\"urich,
           CH-8057, Z\"urich, Switzerland\\
       $^5$Department of Physics, Swiss Federal Institute of Technology,
           ETH H\"onggerberg, CH-8093, Z\"urich, Switzerland}
      

\date{}

\pagerange{\pageref{firstpage}--\pageref{lastpage}}
\pubyear{2000}

\maketitle

\label{firstpage}


\begin{abstract}
  We use the conditional  luminosity function (CLF)  and data from the
  2-degree Field  Galaxy Redshift  Survey  (2dFGRS) to  constrain  the
  average relation between light and mass  in a $\Lambda$CDM cosmology
  with  $\Omega_m=0.23$       and   $\sigma_8=0.74$ (hereafter   WMAP3
  cosmology).  Reproducing the  observed luminosity dependence of  the
  galaxy   two-point   correlation   function   results   in   average
  mass-to-light  ratios  that are  $\sim  35$ percent lower than  in a
  $\Lambda$CDM    cosmology     with $\Omega_m=0.3$ and $\sigma_8=0.9$
  (hereafter WMAP1 cosmology).  This removes an important problem with
  previous  halo occupation models which   had  a tendency to  predict
  cluster mass-to-light  ratios that  were too   high.  For the  WMAP3
  cosmology, our   model yields average  mass-to-light ratios, central
  galaxy luminosities,  halo occupation numbers,  satellite fractions,
  and luminosity-gap  statistics, that are  all in excellent agreement
  with  those obtained  from a 2dFGRS   group catalogue and from other
  independent  studies.   We also use   our  CLF model to  compute the
  probability distribution $P(M  \vert  L_{\rm cen})$, that a  central
  galaxy  of luminosity $L_{\rm  cen}$ resides in a  halo of mass $M$.
  We find this distribution to be  much broader than what is typically
  assumed in halo   occupation  distribution (HOD)  models, which  has
  important  implications  for   the    interpretation  of   satellite
  kinematics and galaxy-galaxy lensing data.  Finally, reproducing the
  luminosity dependence of  the pairwise peculiar velocity dispersions
  (PVDs)  in the 2dFGRS requires  relatively  low mass-to-light ratios
  for  clusters and a  satellite fraction that decreases strongly with
  increasing luminosity.  This is only  marginally consistent with the
  constraints  obtained from the luminosity   dependence of the galaxy
  two-point  correlation  function.  We  argue   that a cosmology with
  parameters   between those of  the  WMAP1  and WMAP3  cosmologies is
  likely to yield results with a higher level of consistency.
\end{abstract}




\section{Introduction}
\label{sec:intro}

Using the observed distribution of galaxies to constrain the cosmology
dependent matter distribution requires  a detailed knowledge of galaxy
bias.  The development of the halo model (see Cooray \& Sheth 2002 for
a detailed review), in which the matter distribution is interpreted in
terms of  its halo building blocks, has  provided us with a convenient
way  to quantify galaxy bias. The  concept is that all galaxies reside
in dark matter  haloes, and that  these haloes themselves are a biased
tracer of the dark matter mass distribution,  the so-called halo bias.
As a  consequence of the hierarchical   nature of structure formation,
more massive haloes are more strongly  clustered (Cole \& Kaiser 1989;
Mo \&   White 1996, 2002), and  the  halo bias  is thus  an increasing
function of halo mass.  Galaxy bias is then  completely specified by a
description of  how galaxies of   different properties are distributed
over dark matter haloes of different masses.

In the  standard halo  occupation models, one  tries to  constrain the
halo occupation distribution (HOD)  $P(N\vert M)$, which expresses the
conditional probability that a halo  of mass $M$ contains $N$ galaxies
(of  a  specified  type).   The  first  moment  of  this  distribution
function,  $\langle  N \rangle_M$,  expresses  the  average number  of
galaxies  as function  of  halo  mass.  Together  with  the halo  bias
$b_h(M)$, this completely  specifies the galaxy bias on  large scales. 
On smaller  scales, however, additional information  is required, such
as  the second  moment of  the  HOD, $\langle  N(N-1) \rangle_M$,  and
information  regarding  the spatial  distribution  of galaxies  within
individual haloes (e.g., Seljak  2000; Scoccimarro \etal 2001; Berlind
\& Weinberg 2002; Cooray \& Sheth 2002; Kang \etal 2002; Berlind \etal
2003).  Additional constraints on the higher moments of $P(N \vert M)$
can be obtained from the  $n$-point correlation functions with $n \geq
3$ (Takada \& Jain 2003; Zheng 2004b).

Numerous studies  have shown  that the observed  two-point correlation
function of galaxies  tightly constrains the  first and second moments
of $P(N\vert M)$  (e.g, Jing, Mo  \&  B\"orner 1998; Peacock \&  Smith
2000; Scranton 2003; Magliochetti \& Porciani 2003; Zehavi \etal 2004,
2005;  Tinker  \etal  2005;  Collister \& Lahav   2005),  and that the
resulting  constraints are  in  good  agreement  with  the  occupation
statistics of dark matter subhaloes (e.g, Kravtsov \etal 2004; van den
Bosch, Tormen \&  Gioccoli 2005c).  The  HOD  modeling has also  been
applied  to  various galaxy  populations at  medium to  high redshifts
(e.g., Bullock, Wechsler \& Somerville 2002; Zheng 2004a; Phleps \etal
2006;  Lee \etal 2006) as  well as to quasars (Porciani, Magliocchetti
\&  Norberg 2004).  Furthermore, Zheng  \& Weinberg  (2005) have shown
that  cosmology and galaxy bias  are not  degenerate.  This means that
one cannot arbitrarily modify the HOD  and fit the observed clustering
of galaxies for  any cosmology; the  HOD technique  can simultaneously
constrain both the galaxy  bias  and cosmology  (see also Zheng  \etal
2002; van den Bosch, Mo \& Yang 2003b; Abazajian \etal 2005).

Since we  know that  galaxy bias is  a function of  galaxy properties,
such as luminosity and color, a natural extension of the HOD modeling
is  to consider  the occupation  statistics  as a  function of  galaxy
properties.  In Yang, Mo \& van den Bosch (2003), we took a first step
in this  direction and introduced the  conditional luminosity function
(hereafter  CLF). The  CLF,  $\Phi(L  \vert M)  {\rm  d}L$, gives  the
average number  of galaxies  with luminosity $L  \pm {\rm  d}L/2$ that
reside  in a  halo of  mass  $M$.  Integrating  the CLF  over a  given
luminosity range  $[L_1,L_2]$, yields  the average number  of galaxies
with $L_1 \leq L \leq L_2$ that reside in a halo of mass $M$:
\begin{equation}
\label{Ntot}
\langle N \rangle_M = \int_{L_1}^{L_2} \Phi(L \vert M) \, {\rm d}L\,,
\end{equation}
Thus, the CLF contains the same information as the first moment of the
halo occupation distribution  $P(N \vert M)$, but it  does so for {\it
  any} luminosity  interval.  In addition, the CLF  also specifies the
total, average luminosity in a halo of mass $M$,
\begin{equation}
\label{ltot}
\langle L \rangle_M =  \int_{0}^{\infty} \Phi(L \vert M) \, L \, {\rm d}L\,,
\end{equation}
and thus specifies the average  relation between light and mass in the
Universe.   As  shown  in  Yang  \etal  (2003),  the  CLF  is  tightly
constrained by  the observed  luminosity function and  the correlation
lengths  of  the galaxy  population  as  function  of luminosity.   In
subsequent  papers, the  CLF has  been  used to  study the  occupation
statistics  as function  of both  luminosity  and galaxy  type at  low
redshift (van  den Bosch, Yang \&  Mo 2003a; Yang  \etal 2005b; Cooray
2005a, 2006)  and high redshift  (Yan, Madgwick \& White  2003; Cooray
2005b,  2006),  to study  the  environment  dependence  of the  galaxy
luminosity  function  (Mo   \etal  2004),  to  constrain  cosmological
parameters (van  den Bosch \etal  2003b; Tinker \etal 2005),  to study
the  pairwise peculiar  velocity  dispersion of  galaxies (Yang  \etal
2004; Jing  \& B\"orner  2004; Li \etal  2006), to  construct detailed
mock  galaxy redshift  surveys (Yang  \etal 2004;  Yan, White  \& Coil
2004; van den Bosch \etal 2005a) and to investigate the luminosity and
type dependence  of the  three-point correlation function  (Wang \etal
2004).  In addition, the CLF  has proven a useful aid for interpreting
the kinematics of  satellite galaxies (van den Bosch  \etal 2004), for
constructing  galaxy  group catalogues  (Yang  \etal  2005a), and  for
furthering our understanding of the galaxy luminosity function (Cooray
\& Milosavljevi\'c 2005).
\begin{figure*}
  \centerline{\psfig{figure=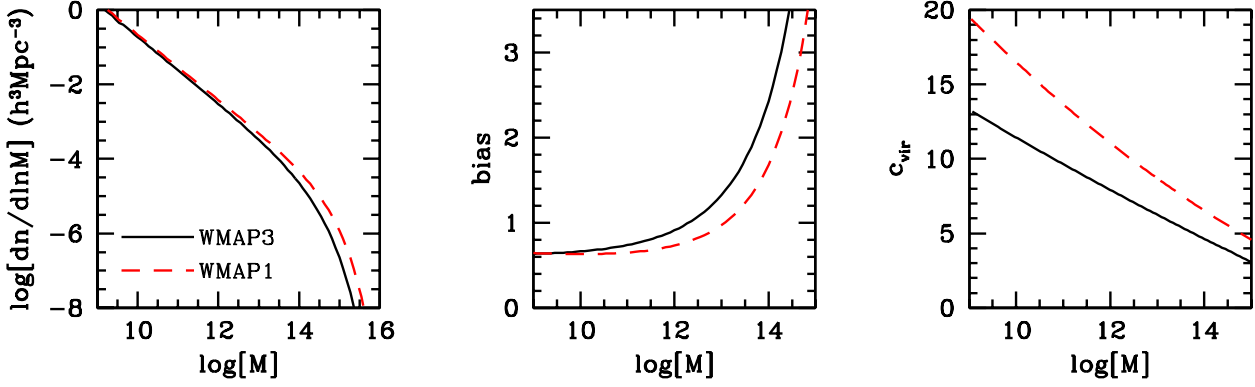,width=0.95\hdsize}}
\caption{Several characteristics of dark matter haloes at $z=0$ in the
  WMAP1 and WMAP3 cosmologies. From  left to right the panels show the
  halo mass functions, the halo bias as function of halo mass, and the
  halo  concentration parameter  $c_{\rm vir}$,  again as  function of
  halo mass.  Note  that in the new WMAP3  cosmology, the abundance of
  massive haloes is strongly  reduced. In addition, dark matter haloes
  are more strongly biased and less strongly concentrated.}
\label{fig:mf}
\end{figure*}

Clearly, the CLF formalism is  a  powerful, statistical tool that  has
many applications.  However,  the occupation statistics  inferred from
the observed  clustering data are  cosmology dependent.  Virtually all
studies  mentioned above have   adopted  a   $\Lambda$CDM  concordance
cosmology   with    a    matter    density   $\Omega_m=0.3$    and   a
Harrison-Zel'dovich,  initial power   spectrum  with a   normalization
$\sigma_8=0.9$.  Recently,  however, the 3-year  CMB  data of the WMAP
mission (Hinshaw \etal 2006; Page \etal 2006) has argued in favor of a
flat  $\Lambda$CDM  cosmology with a  significantly reduced $\Omega_m$
and  $\sigma_8$, and  with a   spectral index   that is  significantly
smaller  than unity (Spergel \etal 2006).   This  has a non-negligible
impact on the halo mass function and the halo bias, both of which play
an important role in the HOD modeling.  The  purpose of this paper is
to revisit some of our conclusions  based on the  CLF formalism in the
revised concordance cosmology.  In  particular, we want to investigate
(i)  whether we  can still simultaneously   fit the galaxy LF  and the
luminosity dependence of the galaxy correlation length, (ii) what this
implies   for the  average  relation  between  light  and  mass in the
universe, and (iii)  whether the  resulting CLF  is consistent  with a
range  of other observations,  including various occupation statistics
inferred from galaxy group  catalogues,  the luminosity dependence  of
the pairwise   peculiar  velocity  dispersions,   and   the  satellite
fractions inferred from  galaxy-galaxy  lensing studies.   Finally, we
improve   upon our previous analysis  by   taking account of the scale
dependence  of  the   halo   bias,   and   by properly   modeling  the
observational data over the light-cone.

This paper is organized  as follows. In \S\ref{sec:theory} we describe
the  theoretical   framework  of   the  CLF  formalism,   including  a
description of  modeling on the  light-cone, and we discuss  how small
changes  in cosmological  parameters impact  on various  statistics of
dark matter haloes.  \S\ref{sec:res} presents  the new CLF for the new
WMAP  concordance cosmology,  and compares  the  mass-to-light ratios,
satellite  fractions, and  the luminosity  gap statistic  predicted by
that  model  to  observational  data.   \S\ref{sec:Nstat}  presents  a
detailed analysis of the halo  occupation numbers predicted by our CLF
model, which are compared to  other HOD models.  In \S\ref{sec:pvd} we
use  mock  galaxy redshift  surveys  to  study  the pairwise  velocity
dispersions of galaxies and  their luminosity dependence. We summarize
our results in \S\ref{sec:concl}.

\section{Theoretical Framework}
\label{sec:theory}

\subsection{Light-Cone Modeling}
\label{sec:lcmod}

As shown in  Yang \etal (2003), the CLF can  be tightly constrained by
fitting it  to the galaxy  luminosity function, $\Phi(L)$, and  to the
galaxy  correlation lengths  as function  of luminosity,  $r_0(L)$.  A
complication arises  from the fact that these  observational data have
been  determined on  a  light-cone.  In  particular,  each data  point
typically derives  from a different light-cone  specified by different
redshift limits  $z_{\rm min}$ and  $z_{\rm max}$.  For example,  in a
flux-limited sample, one has that $z_{\rm max} = z_{\rm max}(L)$.

Since the halo  mass function and the halo bias  are both functions of
redshift,  one  needs  to   properly  integrate  the  model  over  the
light-cone before comparing it to the data. This becomes more and more
important  when $z_{\rm  max}  \gg z_{\rm  min}$.   However, even  for
relatively  nearby surveys,  such  as the  2dFGRS  considered in  this
paper, ignoring this light-cone modeling  may result in errors of 5 to
15  percent. Typically  the errors  are larger  for  brighter samples,
since they cover a larger (deeper) volume.

Within the CLF formalism, the light-cone integrated LF is given by
\begin{equation}
\label{phiLcone}
\Phi(L) = {1 \over V} \int_{z_{\rm min}}^{z_{\rm max}} \rmd z 
{\rmd V \over \rmd  z} \int_0^{\infty} \rmd M \, \Phi(L \vert M,z) \, 
n(M,z) 
\end{equation}
where $\rmd V  /\rmd z$ is the comoving volume  element per unit solid
angle. In  what follows we  will assume that  the CLF does  not evolve
with  redshift, i.e.,  $\Phi(L \vert  M,z) =  \Phi(L \vert  M)$, which
implies that we can write
\begin{equation}
\label{clf}
\Phi(L) = \int_{0}^{\infty} \Phi(L \vert M) \, n_{\rm eff}(M) \, {\rm d}M
\end{equation}
with
\begin{equation}
\label{neff}
n_{\rm eff}(M) \equiv {1 \over V} \int_{z_{\rm min}}^{z_{\rm max}} 
\rmd z {\rmd V \over \rmd z} n(M,z) 
\end{equation}
Note  that this  effective mass  function is  different for  each data
point, i.e., for each different $(z_{\rm min},z_{\rm max})$.

In the case  of the clustering data we proceed as  follows. At a given
redshift, and on large  scales, the two-point correlation function for
dark matter haloes of mass $M$ can be defined as
\begin{equation}
\label{xihh}
\xi_{\rm hh}(r,M,z) = b_h^2(M,r,z) \, \xi_{\rm dm}(r,z)
\end{equation}
with  $b_h(M,r,z)$  the   scale-dependent  halo  bias,  and  $\xi_{\rm
  dm}(r,z)$ the  evolved, non-linear correlation function  of the dark
matter at redshift  $z$.  In what follows we assume  (i) that the mass
dependence of  the halo bias  is separable from the  scale dependence,
and (ii) that the scale  dependent part is independent of redshift. We
thus write that
\begin{equation}
b_h^2(M,r,z) = \tilde{b}_h^2(M,z) \, \zeta(r)
\end{equation}
Using large numerical simulations  Tinker \etal (2005) have shown that
assumption (i) is accurate and that, at $z=0$,
\begin{equation}
\label{radbias}
\zeta(r) = { [1 + 1.17 \, \xi_{\rm dm}(r,0)]^{1.49} \over
[1 + 0.69 \, \xi_{\rm dm}(r,0)]^{2.09}}
\end{equation}
Our assumption  (ii) implies that this  equation also holds  at $z>0$. 
Although this is untested at present, and may well be incorrect, it is
unlikely to have  a significant effect on our  results. After all, the
correction for the scale dependence of the halo bias is only important
for $r  \lta 3 h^{-1} \Mpc$,  which is smaller than  the scales probed
here.  Indeed, if we completely  ignore the scale dependence (i.e., if
we set  $\zeta=1$), we obtain results  that are only  $\sim 5$ percent
different, which is smaller than the measurements errors.

The mass and redshift `dependence of the halo bias can be written as
\begin{eqnarray}
\label{bm}
\tilde{b}_h(M,z) & = & 1 + {1\over\sqrt{a}\delta_c(z)} \
\Bigl[ \sqrt{a}\,(a\nu^2) + \sqrt{a}\,b\,(a\nu^2)^{1-c} - \nonumber \\
& & {(a\nu^2)^c\over (a\nu^2)^c + b\,(1-c)(1-c/2)}\Bigr],
\end{eqnarray}
with $a=0.707$, $b=0.5$, $c=0.6$, and  $\nu = \nu(M,z) = \delta_c(z) /
\sigma(M)$  (Sheth, Mo  \&  Tormen 2001).  Here  $\delta_c(z)$ is  the
critical overdensity required for  spherical collapse at redshift $z$,
and $\sigma(M)$ is the linear  theory rms mass fluctuation on the mass
scale  $M$.   Tinker \etal  (2005),  using state-of-the-art  numerical
simulations, re-investigated the mass  dependence of the halo bias, and
found that equation~(\ref{bm})  accurately fits their simulations, but
with $b=0.35$ and $c=0.80$, which are the values we adopt throughout.
\begin{table*}
\label{tab:cospar}
\caption{Cosmological Parameters.}
\begin{tabular}{lllllllcccccccc}
   \hline
ID & $\Omega_m$ & $\Omega_{\Lambda}$ & $\Omega_b$ & $h$ & $n_s$ & $\sigma_8$ &
$M^*$ & $r_0^{\rm dm}$ &  $a_0$ & $a_1$ & $a_2$ & $c_0$ & $c_1$ & $c_2$ \\
 (1) & (2) & (3) & (4) & (5) & (6) & (7) & (8) & (9) & (10) & 
(11) & (12) & (13) & (14) & (15) \\
\hline\hline
WMAP1 &  $0.30$   & $0.70$ & $0.04$  & $0.70$   & $1.0$   & $0.90$   & 
 $8.57 \times 10^{12}$ & $5.10$ & $-0.056$ & $0.994$ & $-0.001$ & 
 $11.07$ & $-2.49$ & $0.11$ \\
WMAP3 &  $0.238$ & $0.762$ & $0.041$ & $0.734$ & $0.951$ & $0.744$ & 
 $1.36 \times 10^{12}$ & $4.27$ & $-0.078$ & $0.991$ & $-0.002$ & 
 $7.92$ & $-1.70$ & $0.03$ \\
\hline
\end{tabular}
\medskip

\begin{minipage}{\hdsize}
  Parameters of the  two cosmological models discussed in  this paper. 
  Column (1) indicates the name by which we refer to these cosmologies
  in the paper.   Columns (2)-(7) list the matter  density, the energy
  density  associated  with  the  cosmological  constant,  the  baryon
  density,  the   Hubble  parameter,  the  spectral   index,  and  the
  power-spectrum  normalization. Column~(8)  lists  the characteristic
  halo mass  (in $h^{-1} \Msun$), defined  as the mass  scale at which
  the   mass  variance   $\sigma(M)=1.68$.   Column   (9)   lists  the
  correlation  length (in  $h^{-1} \Mpc$)  of the  evolved, non-linear
  matter  field at  $z=0$.  Finally,  columns  (10) to  (15) list  the
  fitting  parameters  that  describe  the relation  between  $M$  and
  $M_{\rm   vir}$    (equation~[\ref{fitMrat}])   and   between   halo
  concentration $c_{\rm vir}$ and $M$ (equation~[\ref{fitcvir}]).
\end{minipage}

\end{table*}

The two-point correlation function for  dark matter haloes of mass $M$
in the  volume of a  light-cone with $z_{\rm  min} \leq z  \leq z_{\rm
  max}$ is given by
\begin{equation}
\label{xihhlc}
\xi_{\rm hh}(r,M) = {\int_{z_{\rm min}}^{z_{\rm max}} \rmd z
\, {\rmd V \over \rmd z} \,n^2(M,z) b_h^2(M,r,z) \xi_{\rm dm}(r,z) \over 
\int_{z_{\rm min}}^{z_{\rm max}} \rmd z \, {\rmd V \over \rmd z} \,
n^2(M,z)}
\end{equation}
(cf. Hamana  \etal 2001). Using  that, on sufficiently  large (linear)
scales, $\xi_{\rm  dm}(r,z) =  D^2(z) \xi_{\rm dm}(r,0)$,  with $D(z)$
 linear growth rate normalized to unity at $z=0$, we obtain that
\begin{equation}
\label{xihhshort}
\xi_{\rm hh}(r,M) = b^2_{h,{\rm eff}}(M) \, \zeta(r) \, \xi_{\rm dm}(r,0)
\end{equation}
with
\begin{equation}
\label{beff}
b^2_{h,{\rm eff}}(M) = {\int_{z_{\rm min}}^{z_{\rm max}} \rmd z \,
{\rmd V \over \rmd z} \, n^2(M,z) \, D^2(z) \, \tilde{b}^2(M,z) \over 
\int_{z_{\rm min}}^{z_{\rm max}} \rmd z \, {\rmd V \over \rmd z} \,
n^2(M,z)}
\end{equation}
Using this effective halo bias, we can write the two-point correlation
function for galaxies of luminosity $L$, on large scales, as
\begin{equation}
\label{galtwopoint}
\xi_{\rm gg}(r,L) = b_{g,{\rm eff}}(L) \, \zeta(r) \, \xi_{\rm dm}(r,0)
\end{equation}
where the effective galaxy bias $b_{g,{\rm eff}}(L)$ is related to the
effective halo  bias, the  effective halo mass  function, and  the CLF
according to
\begin{equation}
b_{g,{\rm eff}}(L) = {1 \over \Phi(L)} \int_0^{\infty} \Phi(L \vert M) \, 
b_{h,{\rm eff}}(M) \, n_{\rm eff}(M) \, \rmd M 
\end{equation}

\subsection{Cosmology}

From  the  above  it is  clear  that the   computation of  the  galaxy
correlation lengths, $r_0(L)$,  defined by $\xi_{\rm  gg}(r_0,L) = 1$,
requires  the  halo mass function, $n(M,z)$,  the  halo bias function,
$\tilde{b}_h(M,z)$,  and    the dark  matter    correlation  function,
$\xi_{\rm  dm}(r,z)$, all of which   are cosmology dependent.  In this
paper we focus on a flat $\Lambda$CDM cosmology  with a matter density
$\Omega_m=0.238$, a baryonic matter density $\Omega_b=0.041$, a Hubble
parameter   $h=H_0/(100 \kmsmpc)=0.734$,  a   power-law initial  power
spectrum  with    spectral  index  $n_s=0.951$    and  a normalization
$\sigma_8=0.744$.  These are the  parameters that best-fit the  3-year
WMAP data (Spergel \etal 2006), and we will refer to this cosmology in
what follows as the WMAP3  cosmology.  For comparison, we also compare
some   of our  results    to   a flat    $\Lambda$CDM  cosmology  with
$\Omega_m=0.3$,    $\Omega_b=0.04$,    $h=0.7$,         $n_s=1.0$  and
$\sigma_8=0.9$.  With strong support from  the first year data release
of  the  WMAP  mission,  this  model  has   has  been  considered  the
concordance `cosmology' in  the vast majority  of all HOD studies.  In
what follows we will refer to a cosmology with these parameters as the
WMAP1 cosmology.

Throughout this paper we compute the halo mass function using the form
suggested by Sheth, Mo \& Tormen (2001), which has been shown to be in
excellent agreement with numerical  simulations as long as halo masses
are defined as the masses  inside a sphere with an average overdensity
of $180$ (Jing  1998; Sheth \& Tormen 1999;  Jenkins \etal 2001; White
2002).  Therefore, in what follows we consistently use that definition
of  halo mass when  referring to  $M$.  The  linear power  spectrum of
density  perturbations  is computed  using  the  transfer function  of
Eisenstein  \& Hu  (1998), which  properly accounts  for  the baryons,
while the evolved, non-linear  power spectrum, required to compute the
dark  matter  correlation  function,  is computed  using  the  fitting
formula of Smith \etal (2003).

The left-hand panel of Fig.~\ref{fig:mf} plots the halo mass functions
at  $z=0$ for  the WMAP1  and WMAP3  cosmologies.  Note  that  the new
concordance cosmology predicts much fewer massive haloes: in fact, the
number density  of haloes  with $M =  10^{15} h^{-1}  \Msun$ ($10^{14}
h^{-1} \Msun$)  is only 19 percent (48  percent) of what it  is in the
WMAP1 cosmology. Clearly, all galaxies assigned to these haloes in the
WMAP1 HOD models  now have to be redistributed  over other haloes. The
middle  panel of  Fig.~\ref{fig:mf} plots  the halo  bias at  $z=0$ as
function of  halo mass.  Although  the overall clustering  strength of
the dark matter is reduced in the WMAP3 cosmology with respect to that
in the WMAP1 cosmology (see Table~1), the halo bias has become larger.
The difference is largest at $M \simeq 3 \times 10^{14} h^{-1} \Msun$,
where  the halo  bias is  $\sim 1.5$  times larger  than in  the WMAP1
cosmology.

As  mentioned above, our  halo masses  $M$ are  defined as  the masses
inside  a  sphere  with  an  average overdensity  of  $180$.   Another
definition of halo mass that  is often adopted is the so-called virial
mass, $M_{\rm vir}$, which indicates  the mass inside a sphere with an
average density equal to $\Delta_{\rm vir}$ times the critical density
for  closure.   The  value  of  $\Delta_{\rm vir}$  follows  from  the
solution to the collapse of a spherical top-hat perturbation under the
assumption that the halo has just virialized, and depends on cosmology
through $\Omega_m(z)$  (Peebles 1980; see  Bryan \& Norman 1998  for a
useful  fitting  function).  Under  the  assumption  that the  density
distribution  of dark  matter  haloes is  well  fit by  a NFW  profile
(Navarro, Frenk \&  White 1997), one can convert  $M$ to $M_{\rm vir}$
(and vice versa) as long as one knows the halo concentration parameter
$c_{\rm vir}$. Using the  $c_{\rm vir}(M_{\rm vir})$ of Macci\`o \etal
(2006), we  find that  the relation between  $M$ and $M_{\rm  vir}$ is
accurately fit (to  better than one percent over  the mass range $10^9
h^{-1}\Msun \leq M \leq 10^{16} h^{-1} \Msun$) by
\begin{equation}
\label{fitMrat}
\log\left[{M_{\rm vir} \over 10^{12} h^{-1} \Msun}\right] = 
a_0 + a_1 y + a_2 y^2
\end{equation}
with  $y =  {\rm log}[M/10^{12}  h^{-1}  \Msun]$. Over  the same  mass
range, the relation between $c_{\rm  vir}$ and $M$ (not $M_{\rm vir}$)
is accurately fit by
\begin{equation}
\label{fitcvir}
c_{\rm vir} = c_0 + c_1 y + c_2 y^2
\end{equation}
The  best-fit parameters  $a_i$ and  $c_i$ ($i=0,1,2$),  for  both the
WMAP1 and the WMAP3 cosmologies, are listed in Table~1.  The resulting
$c_{\rm   vir}(M)$   are   shown    in   the   right-hand   panel   of
Fig.~\ref{fig:mf}:  in  the WMAP3  cosmology  halo concentrations  are
$\sim 30$ percent smaller than  in the WMAP1 cosmology.  This may have
a non-negligible impact on the two-point correlation function on small
scales ($r \lta 50 h^{-1} \kpc$).  In addition, this reduction in halo
concentrations  also has important  implications for  the Tully-Fisher
zeropoint (see  discussions in  van den Bosch  \etal 2003b  and Dutton
\etal 2006).
\begin{table*}
\label{tab:clfparam}
\caption{Conditional luminosity function parameters.}
\begin{tabular}{lccccccccc}
   \hline
ID & $(M/L)_{cl}$ & $(M/L)_0$ & log$M_1$ & log$M_2$ &
$\gamma_1$ & $\gamma_2$ & $\gamma_3$ & $\alpha_{15}$ & $\eta$ \\
 (1) & (2) & (3) & (4) & (5) & (6) & (7) & 
(8) & (9) & (10) \\
\hline\hline
 WMAP1  &  $432^{+92}_{-69}$ & $129^{+26}_{-19}$ & $10.88^{+0.29}_{-0.22}$ & $12.11^{+0.23}_{-0.26}$ & $2.32^{+1.16}_{-0.83}$ & $0.27^{+0.07}_{-0.06}$ & $0.69^{+0.05}_{-0.05}$ & $-1.20^{+0.26}_{-0.26}$ & $-0.25^{+0.09}_{-0.11}$ \\
 WMAP3  &  $350^{+94}_{-67}$  &  $89^{+18}_{-13}$ & $10.70^{+0.25}_{-0.19}$ & $12.02^{+0.25}_{-0.24}$ & $2.96^{+1.71}_{-1.05}$ & $0.27^{+0.07}_{-0.06}$ & $0.70^{+0.06}_{-0.06}$ & $-1.18^{+0.31}_{-0.32}$ & $-0.21^{+0.12}_{-0.12}$ \\
\hline
 WMAP1a &  $434$ & $128$ & $10.90$ & $12.11$ & $2.36$ & $0.27$ & $0.72$ & $-1.08$ & $-0.19$ \\
 WMAP3a &  $363$ &  $89$ & $10.69$ & $12.00$ & $2.74$ & $0.28$ & $0.68$ & $-1.25$ & $-0.24$ \\
\hline
 WMAP3b &  $220$ & $124$ & $10.72$ & $12.33$ & $3.49$ & $0.17$ & $0.66$ & $-1.55$ & $-0.28$ \\
 WMAP3c &  $543$ &  $74$ & $10.57$ & $11.99$ & $3.29$ & $0.34$ & $0.67$ & $-1.38$ & $-0.27$ \\
 WMAP3d &  $214$ & $132$ & $10.62$ & $12.60$ & $6.83$ & $0.15$ & $0.60$ & $-1.88$ & $-0.35$ \\
\hline
\end{tabular}
\medskip

\begin{minipage}{\hdsize}
  Parameters of CLF models.  Column~(1) lists the ID by which we refer
  to  each  CLF  in  the  text.   Columns~(2)  to~(10)  list  the  CLF
  parameters obtained from the  MCMCs.  The upper two entries indicate
  the median  and 68  percent confidence levels,  the middle  two rows
  list the parameters  of the best-fit models, and  the lower two rows
  list   the   parameters  of  three  extreme   models  discussed   in
  \S\ref{sec:pvd}.   Masses and  mass-to-light ratios  are  in $h^{-1}
  \Msun$ and $h \MLsun$, respectively.
\end{minipage}

\end{table*}

\subsection{The Conditional Luminosity Function}
\label{sec:clf}

Following Yang  \etal (2003)  we parameterize the  CLF by  a Schechter
function:
\begin{equation}
\label{phiLM}
\Phi(L  \vert  M)  {\rm  d}L  = {\tilde{\Phi}^{*}  \over  \wLstar}  \,
\left({L \over  \wLstar}\right)^{\walpha} \, \,  {\rm exp}(-L/\wLstar)
\, {\rm d}L,
\end{equation}
where   $\wLstar   =   \wLstar(M)$,   $\walpha   =   \walpha(M)$   and
$\tilde{\Phi}^{*}  = \tilde{\Phi}^{*}(M)$  are all  functions  of halo
mass $M$. We write the average, total mass-to-light ratio of a halo of
mass $M$ as
\begin{equation}
\label{MtoLmodel}
\langle M/L \rangle_M = {1 \over 2} \,
\left({M \over L}\right)_0 \left[ \left({M \over M_1}\right)^{-\gamma_1} +
\left({M \over M_1}\right)^{\gamma_2}\right],
\end{equation}
This parameterization has four  free parameters: a characteristic mass
$M_1$, for  which the mass-to-light  ratio is equal to  $(M/L)_0$, and
two slopes,  $\gamma_1$ and $\gamma_2$,  that specify the  behavior of
$\langle M/L \rangle_M$ at the low and high mass ends, respectively.

A  similar parameterization is used  for the characteristic luminosity
$\wLstar(M)$:
\begin{equation}
\label{LstarM}
{M \over \wLstar(M)} = {1 \over 2} \, \left({M \over L}\right)_0 \,
f(\walpha) \, \left[ \left({M \over M_1}\right)^{-\gamma_1} +
\left({M \over M_2}\right)^{\gamma_3}\right],
\end{equation}
with
\begin{equation}
\label{falpha}
f(\walpha) = {\Gamma(\walpha+2) \over \Gamma(\walpha+1,1)}.
\end{equation}
Here  $\Gamma(x)$   is  the  Gamma  function   and  $\Gamma(a,x)$  the
incomplete Gamma  function.  This parameterization  has two additional
free  parameters: a characteristic  mass $M_2$  and a  power-law slope
$\gamma_3$.   

In our  previous CLF studies, we used  to set $\langle M/L \rangle_M =
(M/L)_{\rm  cl}$ for haloes with $M  \geq 10^{14}  h^{-1} \Msun$, with
$(M/L)_{\rm cl}$     a free  parameter  that describes     the average
mass-to-light ratio of clusters.  This  was motivated  by a number  of
observational studies  (Bahcall, Lubin \&   Norman 1995; Bahcall \etal
2000;   Sanderson \& Ponman 2003),   which indicated that $\langle M/L
\rangle_M$ is roughly constant on the scale  of galaxy clusters.  This
was further supported  by semi-analytical models of galaxy  formation,
which revealed a similar behavior (see Fig.~14 in  van den Bosch \etal
2003a). However,  a number of  studies have suggested that the average
mass-to-light ratio of clusters continues to increase with mass (Adami
\etal 1998; Girardi  \etal 2002; Marinoni \&  Hudson  2002; Bahcall \&
Comerford 2002; Lin, Mohr \& Stanford  2003, 2004; Ramella \etal 2004;
Rines  \etal 2004; Vale \&  Ostriker 2004, 2006;  Popesso \etal 2005).
Especially   the more recent   studies  have  convincingly shown  that
$\langle  M/L \rangle_M  \propto M^{0.2 \pm   0.08}$  on the scale  of
clusters,  virtually independent of  the photometric band in which the
luminosities   are   measured  (Popesso \etal  2005,    and references
therein).   In  this paper,  we therefore   do not  force $\langle M/L
\rangle_M$  to become constant at  large $M$.  Rather we simply adhere
to the functional  form  of equation~(\ref{MtoLmodel}),   according to
which $\langle  M/L \rangle_M \propto M^{\gamma_2}$  at large $M$.  As
we will show below,  this actually yields  values for $\gamma_2$  that
are in excellent  agreement with the  cluster data discussed above. In
order to allow  for a comparison  with our previous models,  we define
$(M/L)_{\rm  cl}$ as  the mass-to-light  ratio  for haloes  with $M  =
10^{14} h^{-1} \Msun$.

For $\walpha(M)$ we adopt a simple linear function of $\log(M)$,
\begin{equation}
\label{alphaM}
\walpha(M) = \alpha_{15} + \eta \, \log(M_{15}),
\end{equation}
with $M_{15}$ the halo mass in units of $10^{15} \msunh$, $\alpha_{15}
= \walpha(M_{15}=1)$, and $\eta$ describes the change of the faint-end
slope  $\walpha$  with  halo  mass.   Note  that  once  $\walpha$  and
$\wLstar$ are  given, the normalization $\tilde{\Phi}^{*}$  of the CLF
is  obtained through equation~(\ref{MtoLmodel}),  using the  fact that
the total (average) luminosity in a halo of mass $M$ is given by
\begin{equation}
\label{meanL}
\langle L \rangle_M = \int_{0}^{\infty}  \Phi(L \vert M) \, L \, {\rm
d}L = \tilde{\Phi}^{*} \, \wLstar \, \Gamma(\walpha+2).
\end{equation}
Finally, we introduce the mass  scale $M_{\rm min}$ below which we set
the CLF to zero; i.e., we assume that no stars form inside haloes with
$M < M_{\rm min}$.  Motivated by reionization considerations (see Yang
\etal 2003 for  details) we adopt $M_{\rm min}  = 10^{9} h^{-1} \Msun$
throughout.  Note,  however,  that  this  lower mass  limit  does  not
significantly influence our results. Changing $M_{\rm min}$ to $10^{8}
h^{-1} \Msun$ or $10^{10} h^{-1}  \Msun$ has only a very modest impact
on the results presented below.

As we will show below, this Schechter-function parameterization of the
CLF  yields good fits  to the  data. In  addition, using  galaxy group
catalogues, Yang  \etal (2005b) has  shown that the Schechter  form is
also consistent  with direct observations  of the CLF. However,  it is
important to keep  an open mind for alternative  functional forms (see
for example  Zheng \etal 2005; Cooray \&  Milosavljevi\'c 2005; Cooray
2005a, 2006).

\subsection{Centrals and Satellites}
\label{sec:censat}

The CLF parameterization presented above does not distinguish a priori
between central and satellite  galaxies. This is somewhat unfortunate,
as there are good reasons to treat these  kinds of galaxies separately
(see  Zheng \etal 2005,   and references  therein).   Although  it  is
straightforward to devise CLF parameterizations with a natural, {\it a
  priori}  split in  central and satellite  galaxies  (e.g., Cooray \&
Milosavljevi\'c   2005;   Cooray   2005a, 2006),    we   stick to  our
Schechter-function parameterization   and apply an  {\it a posteriori}
split into central and satellite components.  In particular, we assume
that the central galaxy is always the brightest galaxy in a halo.

Following  van  den  Bosch  \etal  (2004)  we  write  the  conditional
probability distribution  $P(L_{\rm cen}  \vert M) \rmd  L_{\rm cen}$,
with  $L_{\rm cen}$  the  luminosity  of the  central  galaxy, as  the
product of the CLF and a new function $f(L_{\rm cen},M)$ which depends
on how galaxy luminosities are `drawn' from the CLF:
\begin{equation}
\label{probLbright}
P(L_{\rm cen} \vert M) {\rm d}L_{\rm cen} = 
\Phi(L_{\rm cen} \vert M) \, f(L_{\rm cen},M) \, {\rm d}L_{\rm cen}
\end{equation}
Since the  CLF only  gives the {\it  average} number of  galaxies with
luminosities in  the range $L \pm {\rm  d}L/2$ in a halo  of mass $M$,
there are many different ways  in which one can assign luminosities to
the $N_i$  galaxies of halo  $i$ and yet  be consistent with the  CLF. 
The simplest approach would be  to simply draw $N_i$ luminosities from
$\Phi(L \vert M)$  and to associate $L_{\rm cen}$  with the luminosity
of  the brightest  galaxy.  We  refer to  this luminosity  sampling as
`random',  which results  in  the broadest  $P(L_{\rm  cen} \vert  M)$
possible, at least  when we adhere to the  assumption that the central
galaxy is the brightest halo galaxy.

Alternatively, one could use  a  more constrained approach, and,   for
instance, assume   that the luminosity  of  the  brightest  (and hence
central) galaxy is always larger than $L_1$, defined by
\begin{equation}
\label{Lone}
\int_{L_1}^{\infty} \Phi(L \vert M) \, {\rm d}L \equiv 1
\end{equation}
Although  $L_1 =  L_1(M)$ is  defined  such that  a halo  has {\it  on
  average} one galaxy  with $L \geq L_1$, demanding  that $L_{\rm cen}
\geq L_1$ is equivalent to  assuming that galaxy formation is somewhat
deterministic and that the number  of galaxies with $L \geq L_1(M)$ is
always exactly one.  Hereafter we  shall refer to this sampling method
as `deterministic',  which yields  the narrowest $P(L_{\rm  cen} \vert
M)$  possible.

In the case  of `deterministic' drawing one obviously has that
\begin{equation}
\label{casef}
f(L_{\rm cen},M) = \left\{ \begin{array}{ll}
1 & \mbox{if $L_{\rm cen} \geq L_1(M)$} \\
0 & \mbox{if $L_{\rm cen} < L_1(M)$}
\end{array} \right.
\end{equation}
so that the  expectation value for the luminosity  of a central galaxy
in a halo of mass $M$ is simply given by
\begin{equation}
\label{Lcen}
\langle L_{\rm cen} \rangle_M = 
\int_{L_1}^{\infty} \Phi(L \vert M) \, L \, {\rm d}L = 
\tilde{\Phi}^{*} \, \wLstar \, \Gamma(\walpha+2,L_1/\wLstar)\,.
\end{equation}
In the  case of `random' drawing, one obtains that
\begin{equation}
\label{fran}
f(L_{\rm cen},M) = \left( 1 - {\zeta \over \langle N \rangle_M} \right) \,
{\rm exp}(-\zeta)
\end{equation}
with 
\begin{equation}
\label{zeta}
\zeta = {\langle N \rangle_M - 1 \over \langle N \rangle_M} \,
\int_{L_{\rm cen}}^{\infty} \Phi(L \vert M) {\rm d}L
\end{equation}
(see Appendix~B in van den Bosch \etal 2004 for a derivation), and the
expectation value for $L_{\rm cen}$ has to be computed numerically.

Unless specifically  stated  otherwise, in  what  follows we adopt the
`deterministic' sampling strategy because it allows various statistics
of centrals and satellites  to be computed  analytically from the CLF.
Most of  the results do  not  significantly depend  on this particular
choice. Whenever the detailed form of $f(L_{\rm cen},M)$ is important,
we will present   the results  for  both the   `deterministic' and the
`random' samplings.

\subsection{Parameter Fitting}
\label{sec:mcm}

The CLF,  as specified above,  has a total  of 8 free  parameters: two
characteristic masses; $M_1$ and  $M_2$, four parameters that describe
the various  mass-dependencies $\gamma_1$, $\gamma_2$,  $\gamma_3$ and
$\eta$, a normalization for  the mass-to-light ratio, $(M/L)_0$, and a
normalization of the faint-end slope, $\alpha_{15}$.  The data that we
use to constrain the CLF consists of the 2dFGRS luminosity function of
Madgwick  \etal (2002)  and the  galaxy-galaxy correlation  lengths as
function  of luminosity  obtained  from the  2dFGRS  by Norberg  \etal
(2002).
\begin{figure*}
\centerline{\psfig{figure=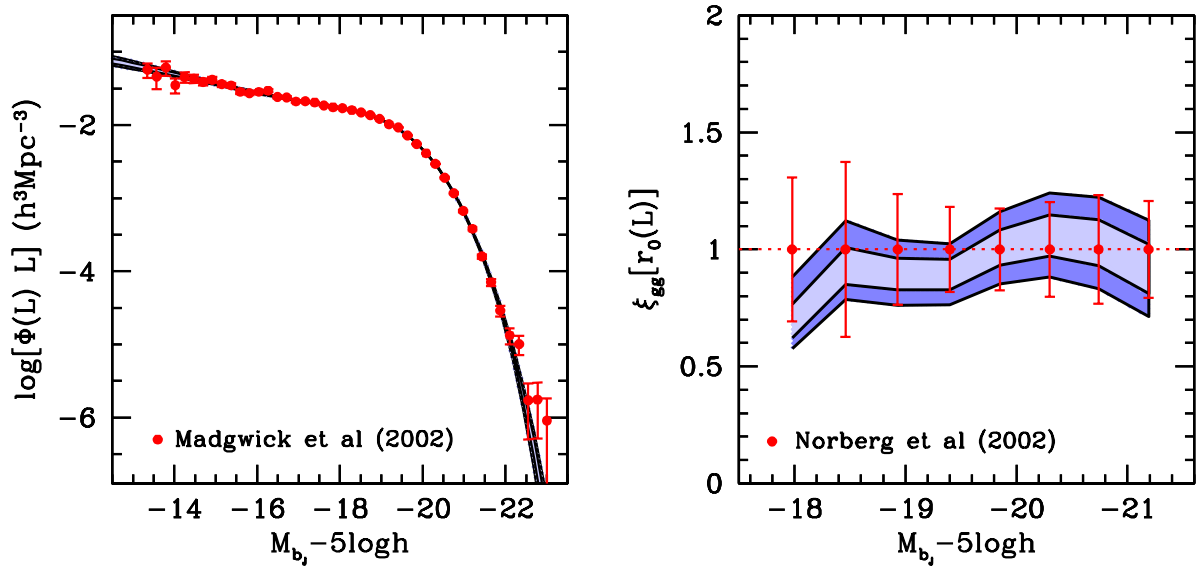,width=0.9\hdsize}}
\caption{The data used to constrain the models (symbols plus
  errorbars), and the  68\% and 95\% confidence limits  from the WMAP3
  MCMC.  The left-hand panel shows the galaxy luminosity function with
  the  2dFGRS data from  Madgwick \etal  (2002), while  the right-hand
  panel shows the values  of the galaxy-galaxy correlation function at
  the correlation lengths of the  magnitude bins used by Norberg \etal
  (2002). For the  data, these are unity by  definition. Note that the
  model accurately fits the data.}
\label{fig:fit}
\end{figure*}

The LF of  Madgwick \etal (2002) has been  determined using the 2dFGRS
data over the  redshift range $0.01 \leq z \leq  0.15$, which we model
using  equations~(\ref{clf})  and~(\ref{neff}),  with $z_{\rm  min}  =
0.01$  and  $z_{\rm  max}  = {\rm  MIN}[0.15,z_{\rm  lim}(L)]$.   Here
$z_{\rm lim}(L)$ is the redshift  at which the apparent magnitude of a
galaxy of  luminosity $L$ is  equal to the  flux limit of  the 2dFGRS,
$b_J = 19.3$ (Colless \etal 2001).

Norberg \etal  (2002) defined a  set of eight volume  limited samples,
each  defined by  two  luminosity limits,  $L_{\rm  min}$ and  $L_{\rm
  max}$, and two redshift limits, $z_{\rm min}$ and $z_{\rm max}$. For
each of these samples,  they determined the correlation length, $r_0$,
defined   by  $\xi_{\rm   gg}(r_0)  =   1$.   We   model   this  using
equation~(\ref{galtwopoint}) with
\begin{equation}
b_{g,{\rm eff}} = {\int_0^{\infty} \langle N \rangle_{M} \, 
b_{h,{\rm eff}}(M) \, n_{\rm eff}(M) \, \rmd M \over \int_0^{\infty} 
\langle N \rangle_{M} \, n_{\rm eff}(M) \, \rmd M}
\end{equation}
where $\langle N  \rangle_M$ is given by (\ref{Ntot}),  but with $L_1$
and $L_2$ replaced by the  luminosity limits $L_{\rm min}$ and $L_{\rm
  max}$ of the volume limited sample under consideration.

To determine the likelihood function of our free parameters we use the
Monte-Carlo  Markov  Chain   (hereafter  MCMC)  technique  (see  e.g.,
Gamerman 1997). Each element of the  chain is a model, consisting of 8
parameters. At any point in the chain we generate a new trial model by
drawing the shifts in the eight free parameters from eight independent
Gaussian  distributions,   centered  on  the  current   value  of  the
corresponding model parameter.  The probability of accepting the trial
model is
\begin{equation}
\label{probaccept}
P_{\rm accept} = \left\{ \begin{array}{ll}
1.0 & \mbox{if $\chi^2_{\rm new} < \chi^2_{\rm old}$} \\
{\rm exp}[-(\chi^2_{\rm new}-\chi^2_{\rm old})/2] & \mbox{if 
$\chi^2_{\rm new} \geq \chi^2_{\rm old}$} \end{array} \right.
\end{equation}
Here  $\chi^2 = \chi^2_{\Phi}  + \chi^2_{r_0}$ with
\begin{equation}
\label{chisqLF}
\chi^2_{\Phi} = \sum_{i=1}^{N_{\Phi}}
\left[ {\Phi(L_i) - \hat{\Phi}(L_i) \over \Delta \hat{\Phi}(L_i)} \right]^2\,,
\end{equation}
and
\begin{equation}
\label{chisqr0}
\chi^2_{r_0} = \sum_{i=1}^{N_{r}}
\left[ {\xi_{\rm gg}(r_{0,i}) - 1 \over 
\Delta \hat{\xi}_{\rm gg}(r_{0,i})} \right]^2,
\end{equation}
where $\hat{.}$ indicates an  observed quantity, and $N_{\Phi}=35$ and
$N_{r}=8$ are the number of data points for the LF and the correlation
lengths,  respectively.   Note  that, by  definition,  $\hat{\xi}_{\rm
  gg}(r_{0,i}) = 1$.

\subsection{The Model}
\label{sec:model}

Using the method described above we construct two chains consisting of
20 million models  each, one for the WMAP1  cosmology and another  for
the WMAP3  cosmology.   Each chain is  thinned by  a  factor $10^4$ to
remove the correlations between neighboring models (see van den Bosch
\etal 2005a for details).  The end result  are two MCMCs consisting of
$2000$ independent models each that properly sample the full posterior
distributions.

Fig~\ref{fig:fit} shows  that the model  based on the  WMAP3 cosmology
accurately fits  the galaxy LF  and the galaxy correlation  lengths as
function  of  luminosity.  The  WMAP1  cosmology,  however, yields  an
equally good  fit to the data (not  shown here, but see  Fig.~3 in van
den  Bosch \etal  2005a).  The  fact  that both  cosmologies allow  an
equally good fit to these  data, despite the large differences in halo
mass function  and halo bias, illustrates that  $\Phi(L)$ and $r_0(L)$
alone allow a  fair amount of freedom in  cosmological parameters (cf. 
van den Bosch \etal 2003b).  However,  as we will see below, the WMAP1
and  WMAP3 cosmologies  predict significantly  different mass-to-light
ratios.
\begin{figure*}
\centerline{\psfig{figure=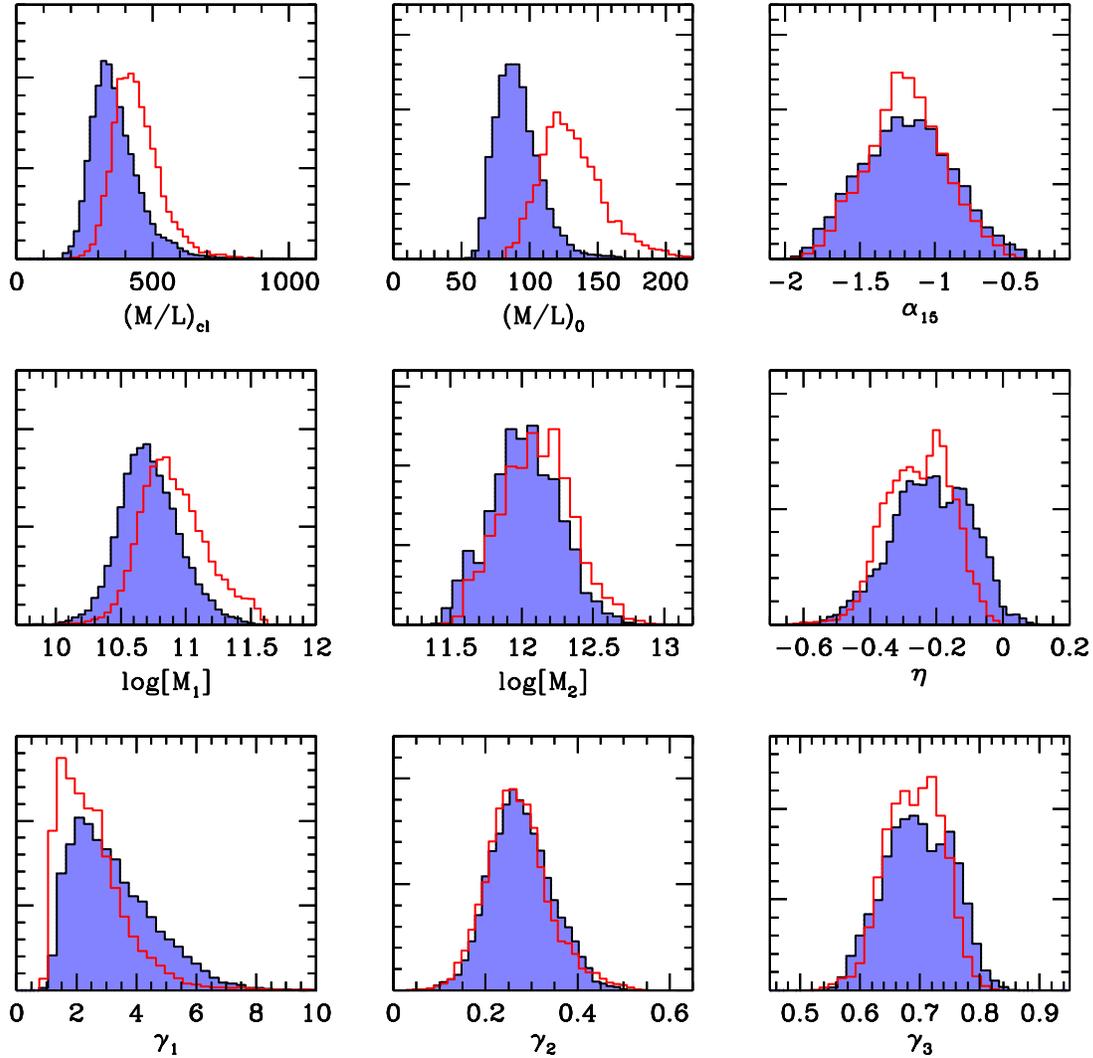,width=0.8\hdsize}}
\caption{Constraints on the nine CLF parameters obtained from our
  MCMCs. The shaded (blue)  and non-shaded (red) histograms correspond
  to the WMAP3 and WMAP1 cosmologies, respectively.  The median and 68
  percent  confidence intervals  of  the distributions  are listed  in
  Table~2.  Masses  and mass-to-light ratios  are in units  of $h^{-1}
  \Msun$ and $h (M/L)_{\odot}$, respectively.}
\label{fig:histo}
\end{figure*}

Fig.~\ref{fig:histo}  plots  the posterior  distributions  of the  CLF
parameters for both the WMAP1 (red, unshaded histograms) and the WMAP3
(blue shaded histograms) MCMCs.   The median and 68 percent confidence
intervals of these distributions  are listed in Table~2.  A comparison
of the  WMAP1 results presented here  with those presented  in van den
Bosch  \etal  (2005a), shows  small  differences  (all  within the  68
percent confidence  levels). These owe to  the fact that (i)  we use a
new model for  the halo bias, including its  scale dependence, (ii) we
properly model  the data over its  light-cone, and (iii)  we no longer
impose  the constraint that  the mass-to-light  ratio is  constant for
haloes with $M \geq 10^{14} h^{-1} \Msun$. 

Comparing  the  WMAP1  and  WMAP3  results,  one  notes  that  several
parameters,  notably  $M_2$,  $\alpha_{15}$,  $\eta$,  $\gamma_2$  and
$\gamma_3$, have virtually the  same likelihood distributions for both
cosmologies.  In  the case of  $(M/L)_{\rm cl}$, $(M/L)_0$  and $M_1$,
however,  the distributions for  the WMAP1  and WMAP3  cosmologies are
clearly offset from each other. As  we show below in more detail, this
mainly reflects  the fact that the mass-to-light  ratios predicted for
the  WMAP3  cosmology  are  significantly  lower than  for  the  WMAP1
cosmology.

\section{Results}
\label{sec:res}

In what  follows we use  the MCMC presented above to  make a number of
predictions  regarding  the  galaxy-dark  matter   connection.   Where
possible, we  will compare these  predictions to  the results obtained
from an analysis of a  large catalogue of  galaxy groups selected from
the 2dFGRS  using the  halo-based  galaxy group finder  of Yang  \etal
(2005a). A short description of  this group catalogue is presented  in
Appendix~A.
\begin{figure*}
\centerline{\psfig{figure=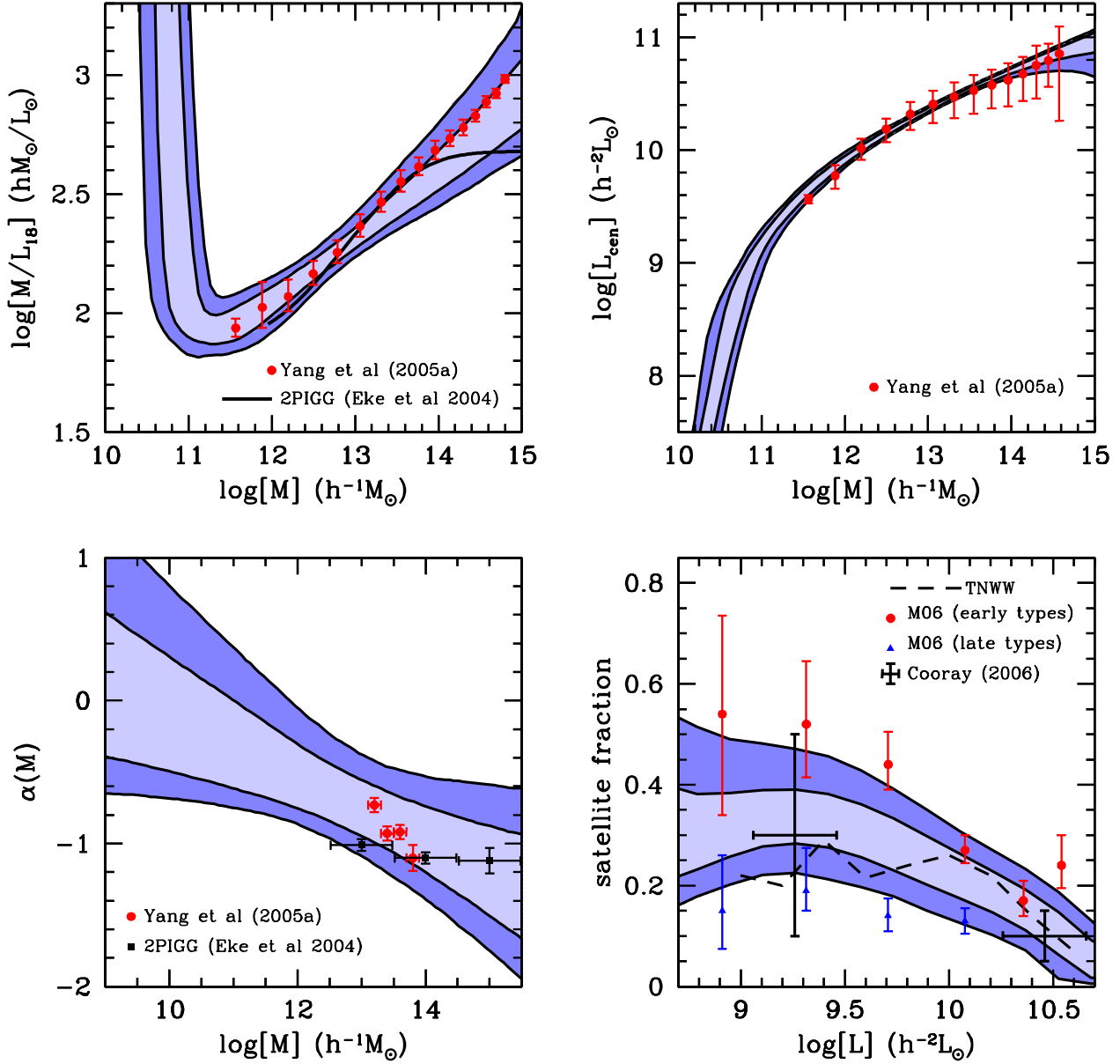,width=0.95\hdsize}}
\caption{Posterior constraints on a number of quantities computed from
  the  WMAP3 MCMC.   The contours  show the  68\% and  95\% confidence
  limits  from the marginalized  distributions.  {\it  Upper left-hand
    panel:} The average ratio between  $M$ and $L_{18}$ as function of
  halo mass. The (red) solid dots indicate the results from our 2dFGRS
  group  catalogue  (see  Table~A1),  while  the  thick  (black)  line
  indicates the  results from the  2PIGG group catalogue of  Eke \etal
  (2004).  {\it Upper right-hand  panel:} The average relation between
  $L_{\rm cen}$  and $M$.   Again, the (red)  solid dots  indicate the
  results  obtained  from  our  2dFGRS group  catalogue.   {\it  Lower
    left-hand panel:}  The faint-end slope  of the CLF,  $\walpha$, as
  function of  halo mass  $M$.  Solid dots  (red) and  squares (black)
  correspond to  the results obtained from our  2dFGRS group catalogue
  and  the  2PIGG  catalogue,  respectively.   {\it  Lower  right-hand
    panel:} The average satellite  fraction as function of luminosity. 
  Solid  circles (red)  and  triangles (blue)  indicate the  satellite
  fractions of early and late type galaxies, respectively, obtained by
  Mandelbaum \etal (2006) from  galaxy-galaxy lensing in the SDSS. The
  thick  dashed  line corresponds  to  the  results  obtained by  TNWW
  (Tinker \etal 2006b) from an HOD analysis of the 2dFGRS, and the two
  thick crosses are the satellite fractions (and their 68\% confidence
  limits) obtained by Cooray (2006) from a CLF analysis of the SDSS}
\label{fig:ml}
\end{figure*}

\subsection{Mass-to-Light ratios}
\label{sec:ml}

As discussed above, the CLF allows one to compute the average relation
between light  and mass in the  Universe. We present  these results in
terms of  the average  mass-to-light ratio as  function of  halo mass,
$\langle M/L_{18} \rangle_M$.  Here $M$  is defined as the mass within
a  radius  inside  of which  the  average  density  is 180  times  the
background  density,  which can  be  converted  into  the virial  mass
$M_{\rm vir}$ using equation~(\ref{fitMrat}). The quantity $L_{18}$ is
the total  luminosity in  the photometric $b_J$  band of  all galaxies
brighter than  $M_{b_J} - 5 \log h  = -18.0$, which is  easy to obtain
from the CLF.

The  upper panels of Fig.~\ref{fig:ml}   show  the 68  and 95  percent
confidence limits on $\langle M/L_{18} \rangle_M$ and $L_{\rm cen}(M)$
as obtained from our WMAP3 MCMC discussed above.  The particular shape
of $\langle M/L_{18} \rangle_M$  holds important information regarding
galaxy formation.  For example, the pronounced minimum  at $M \simeq 2
\times  10^{11} h^{-1}\Msun$ indicates the mass  scale at which galaxy
formation  is most    efficient.  At lower  masses $\langle   M/L_{18}
\rangle_M$ increases dramatically, indicating that galaxy formation is
unable to make galaxies with $M_{b_J} -  5 \log h  \leq -18.0$ in such
low mass haloes.   At the high mass end,  the mass-to-light ratio also
increases,  though  less  rapidly,  indicating that   some  processes,
possibly including AGN feedback, cause galaxy formation to also become
relatively inefficient in massive haloes.

The solid  circles with errorbars  correspond to the  results obtained
from  our 2dFGRS  group  catalogue (Table~A1),  and  are in  excellent
agreement with  the CLF predictions.  It is  extremely reassuring that
two completely different approaches yield average mass-to-light ratios
that are in  such good agreement. Note that  the errorbars indicate the
observed scatter,  not the error  on the mean\footnote{Since  the halo
  masses are  estimated from the group  luminosities (see Appendix~A),
  this scatter is a lower limit on the true scatter.}.

According  to   our parameterization,  at  the high  mass end $\langle
M/L_{18} \rangle_M \propto  M^{\gamma_2}$.  We obtain that $\gamma_2 =
0.27^{+0.07}_{-0.06}$ (see Table~2), in good agreement with our 2dFGRS
galaxy group results and with a wide  range of additional studies (see
\S\ref{sec:clf}).  The  solid  (black) line,   however,  indicates the
mass-to-light ratios  obtained by Eke \etal  (2004) from  their 2dFGRS
group  catalogue  called 2PIGG.  Although the  agreement  with our CLF
predictions and with  the results from our  2dFGRS  group catalogue is
good   for $M \lta 10^{14}   h^{-1} \Msun$, the 2PIGG-catalogue yields
that $\gamma_2  \rightarrow  0$  for  $M \gta  10^{14}  h^{-1} \Msun$.
Similar   results were obtained  by Bahcall,  Lubin  \& Norman (1995),
Bahcall \etal (2000) and Sanderson \&  Ponman (2003).  To test whether
the clustering data itself  can discriminate between these results, we
have constructed a  CLF MCMC in which  we  set $\langle M/L \rangle  =
(M/L)_{\rm cl}$ for $M \geq 10^{14} h^{-1} \Msun$. The resulting model
can fit the observed $\Phi(L)$ and  $r_0(L)$ equally well as the model
presented here,   indicating that the   clustering data alone  can not
meaningfully constrain  the slope  of  the relation  between mass  and
light  on the  scale  of clusters.   The  simple  fact that  two group
catalogues  constructed  from the    same  data  set  (2dFGRS)   yield
predictions  that are very  different,  accentuates the need for  more
thorough investigations.

A comparison  with Fig.~3 in van  den Bosch \etal  (2005a), shows that
the  mass-to-light   ratios  predicted   by  the  CLF   formalism  are
significantly  lower in  the WMAP3  cosmology, compared  to  the WMAP1
cosmology.  This  difference is most  pronounced near the  minimum ($M
\simeq  3 \times  10^{11}  h^{-1} \Msun$,  where  the WMAP1  cosmology
predicts mass-to-light  ratios that are  $\sim 45$ percent  higher. At
the massive end the difference is less pronounced, but at $M = 10^{14}
h^{-1}  \Msun$ the  WMAP1  mass-to-light ratios  are  still $\sim  25$
percent  higher than  for the  WMAP3  cosmology. The  reason for  this
change is a  rather complicated mix of effects.  First  of all, in the
WMAP3  cosmology  there  are  much fewer  massive  haloes.   Secondly,
changing the cosmology  from WMAP1 to WMAP3 decreases  the dark matter
correlation  length  (see Table~1).   Consequently,  galaxies have  to
become  more  strongly  biased   in  order  to  match  their  observed
correlation lengths.   To some extent, this  is automatically achieved
by the fact that  the halo bias is larger in the  WMAP3 cosmology (cf. 
Fig.~\ref{fig:mf}).   However, since  the strength  of this  effect is
strongly dependent  on halo  mass, it is  difficult to  make intuitive
predictions.  Our  analysis shows that  all these effects  conspire to
cause the average mass-to-light ratios to decrease on all mass scales.

The fact that the average mass-to-light ratios are cosmology dependent
is   of great importance.  As  we have shown   in  van den Bosch \etal
(2003b, hereafter  BMY03), it  allows  us to  put tight  constraints on
cosmological parameters, in particular  on $\Omega_m$  and $\sigma_8$.
In principle, there is  a wide range of cosmologies  that allow one to
accurately fit  both $\Phi(L)$  and  $r_0(L)$. Changing  the cosmology
typically implies a    change in the halo   bias  and in  the  overall
clustering strength of the  dark matter.  In order  to maintain a good
fit to   the observed $r_0(L)$ one  has  to redistribute galaxies over
haloes of  different masses in  order to counterbalance the changes in
$b_h(M)$ and $\xi_{\rm dm}(r)$.  As long as these  changes are not too
large, one can  always find a characteristic  halo  mass that has  the
right bias so that $r_0(L)$ is consistent with the data.  However, all
these  different models will  predict  different mass-to-light  ratios
$\langle M/L \rangle_M$,  simply  because they require  different halo
occupation   statistics.  Therefore,  any  constraints on  the average
mass-to-light ratios of  dark matter haloes, on  any mass scale,  will
dramatically tighten the constraints on cosmological parameters.

One such constraint   comes  from galaxy  clusters.  Numerous studies,
based on different  techniques, have measured the mass-to-light ratios
of clusters of galaxies (e.g,  Carlberg \etal 1996 Bahcall \etal 2000;
Lin,  Mohr \& Stanford  2004; Popesso \etal  2005). As shown in BMY03,
all these measurements  are  in good agreement   with each other   and
suggest that $\langle M_{\rm vir}/L_B \rangle_{\rm  cl} = (350 \pm 70)
h \MLsun$.   Using this as  a constraint on the  CLF models puts tight
constraints on the cosmological parameters. In fact, combining the CLF
analysis  with  the  first  year  WMAP results,   BMY03 obtained  that
$\Omega_m = 0.25^{+0.10}_{-0.07}$ and $\sigma_8 = 0.78 \pm 0.12$ (both
95\% CL), in excellent agreement with the 3-year results from the WMAP
mission  (Spergel  \etal  2006).   The  main problem   with  the WMAP1
cosmology  is that it  predicts mass-to-light ratios for clusters that
are too large  (see also  Tinker  \etal 2005; Vale  \& Ostriker 2006).
The  good agreement, both among  these different  studies and with the
latest CMB constraints, demonstrates the  strength and reliability  of
the CLF formalism (or equivalent techniques), especially when combined
with  constraints on mass-to-light   ratios.  It indicates  that  halo
occupation modeling has matured to the point  where it can be used to
obtain tight and  reliable constraints on cosmological parameters (see
also Zheng \& Weinberg 2005).

\subsection{Faint End Slope of CLF}
\label{sec:alpha}

The  lower left-hand  panel of  Fig.~\ref{fig:ml} shows  that  our CLF
model  predicts  that  the  faint-end  slope of  the  CLF,  $\walpha$,
decreases with increasing  halo mass. At around the  cluster scale the
models favor  fairly steep faint-end slopes with  $\walpha \simeq -1.2
\pm 0.3$ (68\% CL), in  good agreement with independent studies of the
luminosity functions  of individual clusters  (e.g., Sandage, Bingelli
\& Tammann  1985; Beijersbergen \etal 2002; Trentham  \& Hodgkin 2002;
Trentham \&  Tully 2002). Note,  however, that since the  CLF reflects
the {\it average}  luminosity function for haloes of  a given mass, it
is not necessarily  a good description of the  luminosity functions in
{\it individual}  systems.  A more meaningful  comparison is therefore
with  the  CLFs  that  one  can  obtain  directly  from  galaxy  group
catalogues, by combining  all groups in a relatively  narrow mass bin. 

The  solid  squares indicate  the  faint-end  slope  of the  CLF  thus
obtained by Eke \etal (2004) from the 2PIGG  group catalogue (see also
Robotham \etal 2006),  while the  (red) solid  dots  correspond to the
results obtained   from  our 2dFGRS group    catalogue (see Yang \etal
2005b). Both  results are in reasonable  agreement with each other and
with the confidence limits obtained from  our CLF analysis (though the
latter are not  very strict). Note that our  results only extent to $M
\lta 10^{14} h^{-1}  \Msun$; in order  to obtain a  sufficiently large
number of massive  groups, one needs to  probe out to relatively  high
redshifts.  However, because  of the flux-limit of  the  2dFGRS we can
not probe the CLF of these groups  to sufficiently low luminosities to
be able to extract a  reliable measure of the  faint-end slope. 

\subsection{Satellite Fractions}
\label{sec:satfrac}

The satellite fraction as function of luminosity, $f_{\rm sat}(L)$, is
an important quantity  for  a proper interpretation  of  galaxy-galaxy
lensing measurements  (Guzik  \& Seljak 2002;  Mandelbaum  \etal 2006;
Yang \etal 2006) and  for understanding pairwise  velocity dispersions
(Slosar, Seljak   \&  Tasitsiomi  2006;  see   also  \S\ref{sec:pvd}).
Therefore, it is useful to check what our  CLF models predict in terms
of $f_{\rm sat}(L)$.

A satellite galaxy  most likely resides in a more  massive halo than a
central galaxy  of the  same luminosity.  Since  halo bias  depends on
halo  mass, the large  scale bias  of galaxies  of a  given luminosity
depends   strongly  on  what   fraction  of   them  are   satellites.  
Consequently,  the   observed  clustering  strength   as  function  of
luminosity  can put  strong constraints  on $f_{\rm  sat}(L)$.  In the
`deterministic' case (see  \S\ref{sec:censat}), the satellite fraction
as function of luminosity is given by
\begin{equation}
\label{fsat}
f_{\rm sat}(L) = {1 \over \Phi(L)} \int_{M_{\rm one}}^{\infty} 
\Phi(L \vert M) \, n(M) \, {\rm d}M
\end{equation}
with $M_{\rm  one}$ defined according  to $L_1(M_{\rm one}) =  L$.  In
words, $M_{\rm  one}$ is the mass  scale at which one  has exactly one
galaxy   brighter   than   $L$,   which  is   easily   computed   from
eq.~(\ref{Lone}) using  a root finder.  The lower  right-hand panel of
Fig.~\ref{fig:ml}  shows the 68  and 95  percent confidence  levels on
$f_{\rm sat}(L)$  computed from our CLF MCMC.   The satellite fraction
decreases  with increasing  luminosity,  from $32  \pm  6$ percent  at
$M_{b_J}-5{\rm  log}h=-17$ to  $11  \pm 4$  percent at  $M_{b_J}-5{\rm
  log}h=-21$  (both 68\% CL).   In the  WMAP1 cosmology  the satellite
fractions  are  about $+5$  percent  higher,  which  is a  small  (but
systematic) difference compared to the model uncertainties for a given
cosmology.
\begin{figure*}
\centerline{\psfig{figure=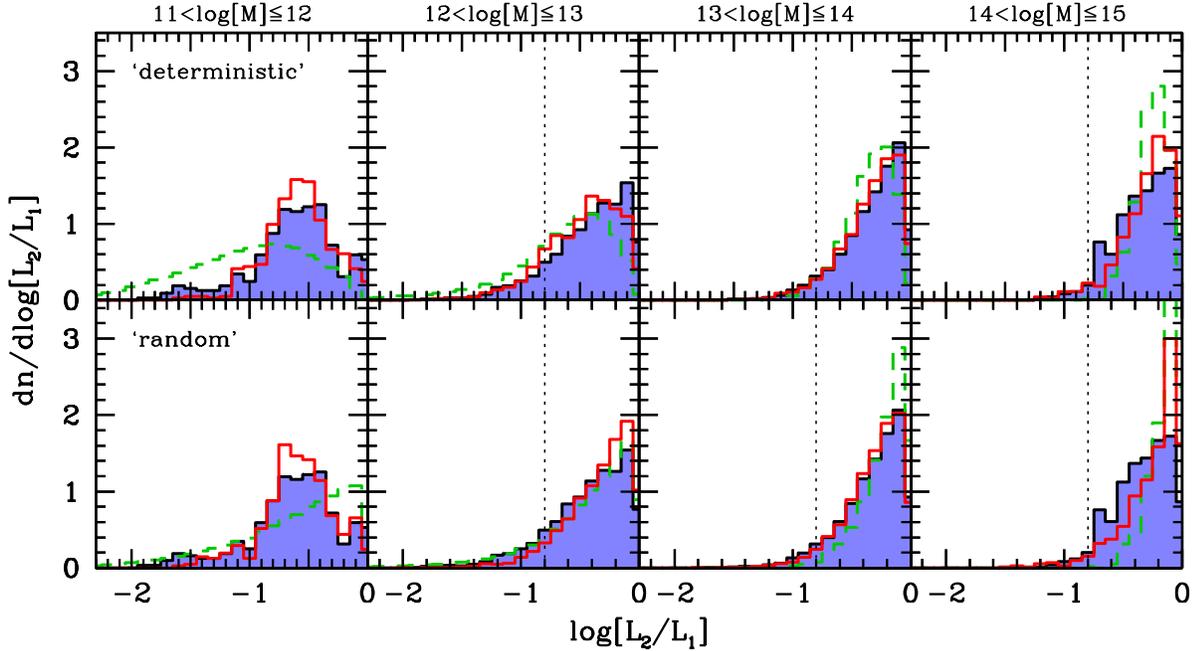,width=0.9\hdsize}}
\caption{The luminosity-gap statistic: The (blue) shaded histograms
  indicate  the  distribution  of  $\log[L_2/L_1]$ obtained  from  our
  2dFGRS group catalogue for four different bins in halo (group) mass,
  as indicated (all masses are  in $h^{-1} \Msun$).  Here $L_i$ is the
  luminosity of the $i^{\rm th}$ brightest galaxy in a given group (or
  halo). The dashed (green)  histograms show the results obtained from
  our best-fit CLF model, when using either the `deterministic' (upper
  panels)  or `random'  (lower  panels) sampling  strategy. The  solid
  (red)  histograms   show  the  results  obtained   from  mock  group
  catalogues constructed from these CLF models.  These can be compared
  directly to the 2dFGRS results.   Note that the overall agreement is
  very satisfactory, both for the `deterministic' and for the `random'
  sampling   strategy.  The  vertical   dashed  lines   correspond  to
  $\log[L_2/L_1]=-0.8$ and  mark the point where  the second brightest
  galaxy is exactly two magnitudes  fainter than the brightest galaxy. 
  Groups  to  the left  of  this line  are  sometimes  referred to  as
  ``fossil groups''.}
\label{fig:gap}
\end{figure*}

For comparison, we overplot the  results from three different studies. 
The  dashed line shows  the satellite  fractions corresponding  to the
fiducial HOD model of Tinker \etal (2006b, hereafter TNWW). This model
is constrained by the luminosity dependence of the clustering strength
in  the 2dFGRS.   The  two crosses  indicate  the satellite  fractions
inferred by Cooray (2006) from a CLF analysis of the Sloan Digital Sky
Survey (hereafter  SDSS), with vertical errorbars  indicating the 68\%
confidence  levels.  Finally,  the  solid circles  and triangles  with
vertical  errorbars  (68\% CL)  indicate  the  satellite fractions  of
early-  and late-type galaxies,  respectively, inferred  by Mandelbaum
\etal   (2006)  from   a   galaxy-galaxy  lensing   analysis  of   the
SDSS\footnote{For the  results of  Mandelbaum \etal (2006)  and Cooray
  (2006) we  have  converted the  SDSS  $r$-band  luminosities to  the
  $b_J$-band  using  the simplifying  assumption  that the  luminosity
  ratio  $L/L^{*}$   is  independent   of  the  photometric   band.}.  
Remarkably, all  these results are  in good agreement with  each other
and with our CLF constraints.

\subsection{Fossil Groups and the Luminosity-Gap Statistic}
\label{sec:gap}

Another useful statistic is the ratio $L_2/L_1$ of the luminosities of
the  second brightest  and brightest  galaxies  in a  given halo.   As
discussed in  D'Onghia \etal  (2005) and Milosavljevi\'c  \etal (2006)
this statistic quantifies  the dynamical age of a  system of galaxies:
haloes  with $L_2/L_1$  close to  unity must  be relatively  young, as
dynamical friction  will cause multiple luminous galaxies  in the same
halo to merge on a relatively short time scale.

In  Fig.~\ref{fig:gap}  we  compare  this  `luminosity-gap'  statistic
obtained  from  our  2dFGRS  group catalogue  (shaded  histograms)  to
results obtained  from our  CLF. To that  extent we populate  the dark
matter haloes in a $300  h^{-1} \Mpc$ cosmological simulation box (see
Appendix~B) with galaxies according  to our best-fit CLF model (called
WMAP3a  in Table~2) using  both the  `deterministic' and  the `random'
formalisms described  in \S\ref{sec:censat}. The results  are shown as
the  dashed  (green)  histograms   in  the  upper  and  lower  panels,
respectively. Both models clearly  predict that the average luminosity
gap increases with decreasing halo mass, in qualitative agreement with
the data.   While the `deterministic'  model predicts that there  is a
deficit of low  mass haloes with $L_2/L_1 \sim  1$, the `random' model
always  predicts  distributions  of  $L_2/L_1$  that peak  at  unity.  
Especially for  haloes with $12 < \log[M/(h^{-1}\Msun)]  \leq 13$, the
latter seems  to be in better  agreement with the  data.  However, the
comparison is not entirely fair. After all, the data has been obtained
from a group catalogue, which suffers from interlopers, incompleteness
and errors in halo mass. 

We  therefore use  the populated  simulation box  to construct  a mock
galaxy redshift survey (as described in detail in Appendix~B) to which
we  apply the group  finder of  Yang \etal  (2005a).  The  solid (red)
histograms  in Fig.~\ref{fig:gap}  show  the luminosity-gap  statistic
obtained  from these  mock group  catalogues.  Overall,  the agreement
between the model and the data is remarkably good, indicating that our
CLF   model  predicts   realistic  values   of   $L_2/L_1$.   Somewhat
unfortunately,  the   differences  between  the   `deterministic'  and
`random'  models are  now much  suppressed, so  that it  is  no longer
possible to clearly discriminate  between these two models.  Note that
the mock  group catalogue yields a distribution  of $\log[L_2/L_1]$ in
the  lowest  halo mass  bin  which is  very  different  from the  true
underlying  distributions (dashed  histograms), indicating  that these
low-mass   haloes  suffer   quite   substantially  from   interlopers,
incompleteness effects and  errors in halo mass. For  the more massive
haloes,  however, the  agreement  between the  true distributions  and
those obtained from the mock group catalogue is very satisfactory.

Systems with a relatively large luminosity gap, which most likely owes
to the fact  that the brightest  galaxies in the  halo have merged, are
often termed ``fossil groups''  and have received a significant amount
of attention in the recent literature (see Vikhlinin \etal 1999; Jones
\etal 2003; D'Onghia  \etal   2005; Milosavljevi\'c \etal   2006;  and
references therein).  Following Jones \etal (2003) and Milosavljevi\'c
\etal (2006) we define systems in which the second brightest galaxy is
at least   2  magnitudes  fainter than   the brightest  galaxy  (i.e.,
$\log[L_2/L_1]\leq  -0.8$,   indicated as  dotted   vertical  lines in
Fig.~\ref{fig:gap})  as  ``fossil''  systems.   With our 2dFGRS  group
catalogue, we are  in a unique  position to determine  the fraction of
fossil groups from a  large and complete  sample of optically selected
galaxy groups. We obtain that the fraction of fossil systems increases
with decreasing halo mass  from $3.6$ percent  for  groups with $14  <
\log[M/(h^{-1} \Msun)] \leq 15$ to $6.5$ percent for groups with $13 <
\log[M/(h^{-1} \Msun)] \leq 14$ to $13.4$ percent  for groups with $12
< \log[M/(h^{-1} \Msun)]  \leq 13$ (in all  three cases the Poissonian
errors  are less than  $0.1$  percent).   For comparison, Jones  \etal
(2003)  obtained an incidence rate of  $8$ to $20$ percent for systems
with an X-ray luminosity from diffuse, hot gas of $L_{\rm X, bol} \geq
2.5  \times 10^{41} h^{-2}  {\rm erg}{\rm  s}^{-1}$.  Although this is
relatively high  compared to the fossil fractions  in our 2dFGRS group
catalogue, the latter have not been X-ray selected which complicates a
straightforward comparison.  In  a recent paper, D'Onghia \etal (2005)
used  detailed hydrodynamical simulations  to  predict the fraction of
haloes with $M \sim 10^{14} h^{-1} \Msun$ that have $\log[L_2/L_1]\leq
-0.8$.  From a total of twelve  simulated groups, they obtain a fossil
fraction  of $33 \pm  16$  percent.  Although consistent  with ours at
their $2 \sigma$  level, the much lower  fraction of fossil systems in
our 2dFGRS group  catalogue suggests a potential over-merging  problem
in their simulations.
\begin{figure*}
\centerline{\psfig{figure=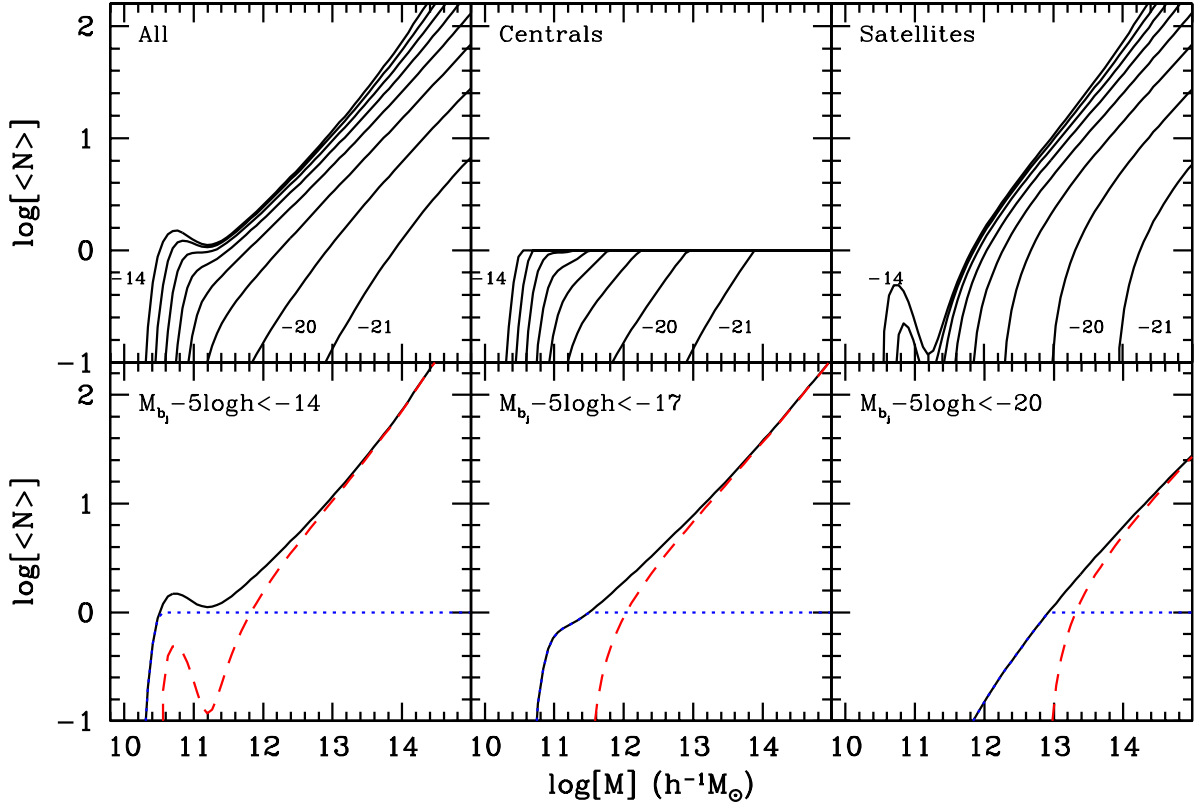,width=0.9\hdsize}}
\caption{Occupation statistics of our best-fit CLF model in the WMAP3 
  cosmology.   The upper panels  plot the  average number  of galaxies
  brighter than  a certain magnitude limit  as function of  halo mass. 
  From left to right the  upper panels show the occupation numbers for
  all galaxies,  for central galaxies, and for  satellite galaxies. In
  each panel,  the different  lines correspond to  different magnitude
  limits:  from   left  to  right,  $M_{b_J}   -  5  {\rm   log}  h  =
  -14,-15,...,-21$.   The lower  panels show  the  combined occupation
  numbers for  three different  magnitude limits, as  indicated. These
  figures  illustrate  that  the  functional form  of  the  occupation
  numbers  predicted by our  CLF models  are in  qualitative agreement
  with   HOD   models,    with   numerical   simulations,   and   with
  semi-analytical models for galaxy  formation, but with one important
  difference:  the zero-to-unity  transition of  $\langle  N_{\rm cen}
  \rangle_M$  is significantly broader  than in  most HOD  models (see
  text for detailed discussion).}
\label{fig:Nstat}
\end{figure*}

\section{Halo Occupation Statistics}
\label{sec:Nstat}

In  this section we   describe the link  between the  CLF and the more
often used HOD models.  The latter aim at constraining the conditional
probability   distribution $P(N \vert   M)$ that  a   halo of mass $M$
contains $N$  galaxies. Here, and  in  what follows, whenever we  talk
about the occupation numbers $N$, we mean  the number of galaxies {\it
  brighter than a given luminosity limit} $L_{\rm min}$.  Most studies
to date have  only  focused on the  first  moment of $P(N  \vert M)$,
which specifies the mean occupation  numbers as function of halo mass.
The same information can be extracted trivially from  the CLF, for any
$L_{\rm min}$, so that the relation between $P(N \vert M)$ and $\Phi(L
\vert M)$ is given by
\begin{equation}
\label{meannm}
\langle N \rangle_M = \sum_{N=0}^{\infty} N \,  P(N \vert M) =
\int_{L_{\rm min}}^{\infty} \Phi(L \vert M) \rmd L
\end{equation}
It is  interesting to compare the  shape of this $\langle N \rangle_M$
{\it  predicted} by the  CLF, with  the  shape that is  typically {\it
  assumed}   in HOD models.   Early  HOD   models  often assumed  that
$\langle N \rangle_M$ follows a simple power-law (Jing, Mo \& B\"orner
1998;   Seljak 2000; Scoccimarro  \etal  2001;   Scranton 2002;  Rozo,
Dodelson  \& Frieman 2004;   Collister  \& Lahav   2005) or  a  broken
power-law (Berlind \& Weinberg 2002;  Magliocchetti \& Porciani 2003).
More recently,    it has become   practice  to adopt a   somewhat more
complicated form,  motivated  by a separate  treatment  of central and
satellite galaxies, i.e.,  $\langle N \rangle_M  = \langle N_{\rm cen}
\rangle_M  + \langle N_{\rm   sat}  \rangle_M$ (e.g., Abazajian  \etal
2005; Sefusatti \& Scoccimarro   2005; Zehavi \etal 2004;  Zheng \etal
2005; Tinker, Weinberg \& Zheng 2006a).  In  all these models $\langle
N_{\rm sat} \rangle_M$ is modeled  as  a power-law (sometimes with  a
break at  small  $M$),  while   $\langle N_{\rm cen}    \rangle_M$  is
considered to change from zero at low $M$ to unity at high $M$, either
via a simple step function at a characteristic mass, or via a somewhat
broader  transition   function.   These functional  forms  are largely
motivated by the occupation statistics of dark matter subhaloes and of
galaxies in  numerical     simulations and  semi-analytical     models
(Kauffmann  \etal 1999;  Benson \etal 2000;   Sheth \& Diaferio  2001;
Berlind \etal 2003; Kravtsov \etal 2004; Zheng \etal 2005).

In  the  case  of our  CLF,  no  assumptions  are made  regarding  the
functional forms of either $\langle N_{\rm cen} \rangle_M$ or $\langle
N_{\rm sat} \rangle_M$.  In  fact, we split the Schechter-function CLF
a  posteriori in contributions  from central  and satellite  galaxies. 
Following the `deterministic'  method described in \S\ref{sec:censat},
the  occupation statistics  of central  and satellite  galaxies follow
trivially from $\langle N \rangle$: if $\langle N \rangle \geq 1$ than
$\langle N_{\rm cen}  \rangle = 1$ and $\langle  N_{\rm sat} \rangle =
\langle N \rangle - 1$. On the  other hand, if $\langle N \rangle < 1$
than $\langle  N_{\rm cen} \rangle  = \langle N \rangle$  and $\langle
N_{\rm sat} \rangle = 0$.

With  one  additional assumption, one   can  in fact  derive the  full
probability distribution $P(N \vert M)$ from the CLF. Motivated by the
fact that dark matter subhaloes reveal Poissonian statistics (Kravtsov
\etal 2004), it  has become  standard  to assume  that  the number  of
satellite galaxies follows a Poisson distribution.   If we follow this
assumption, which  is also the   standard procedure in HOD models,  we
have that
\begin{equation}
\label{pnm}
P(N \vert M) = \left\{ 
\begin{array}{ll}
P_1(N \vert M) & \mbox{if $M > M_{\rm one}$} \\
P_2(N \vert M) & \mbox{otherwise}
\end{array}\right.
\end{equation}
Here $M_{\rm one}$ is defined in \S\ref{sec:satfrac},
\begin{equation}
\label{pone}
P_1(N \vert M) = \left\{ 
\begin{array}{ll}
0 & \mbox{if $N = 0$} \\
{\langle N_{\rm sat} \rangle^{N-1} \over (N-1)!} 
\exp\left[{-\langle N_{\rm sat} \rangle}\right]  & \mbox{otherwise}
\end{array}\right.
\end{equation}
and
\begin{equation}
\label{ptwo}
P_2(N \vert M) = \left\{ 
\begin{array}{lll}
1 - \langle N_{\rm cen} \rangle & \mbox{if $N = 0$} \\
\langle N_{\rm cen} \rangle     & \mbox{if $N = 1$} \\
0                               & \mbox{otherwise} 
\end{array}\right.
\end{equation}
\begin{figure*}
\centerline{\psfig{figure=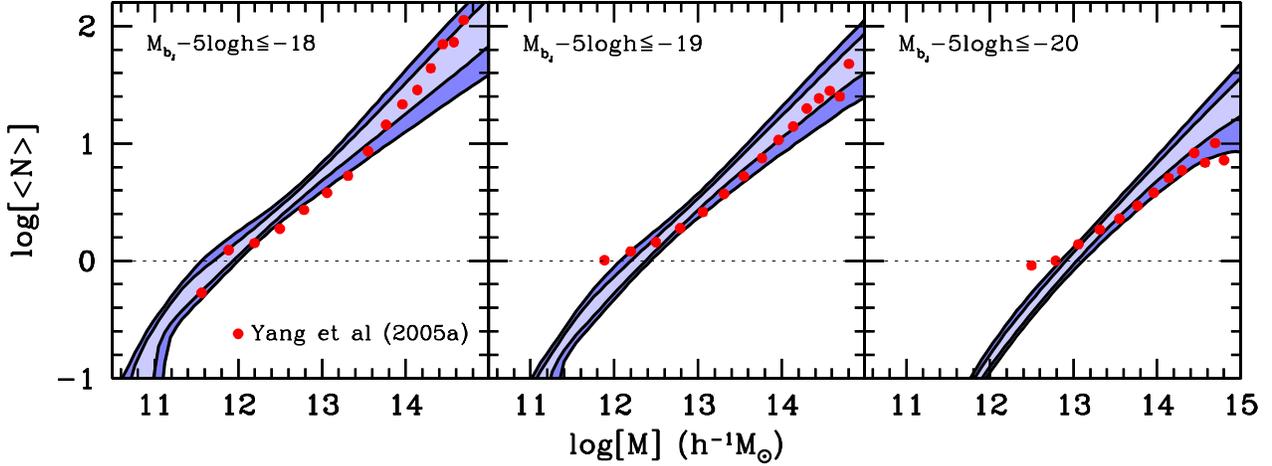,width=0.95\hdsize}}
\caption{The contours  show the  68\% and  95\% confidence
  limits on the halo occupation  numbers obtained from our WMAP3 MCMC. 
  Results  are   shown  for  three  different   magnitude  limits,  as
  indicated, and are compared to  the results obtained from our 2dFGRS
  group catalogue (red, solid dots, see Table~A1).}
\label{fig:honfit}
\end{figure*}

The  upper left-hand panel  of Fig.~\ref{fig:Nstat}  plots the average
number of  galaxies brighter than a given  magnitude limit as function
of halo mass.  The magnitude limits are, from  left to right, $M_{b_J}
- 5 {\rm  log} h = -14,-15,...,-21$.  Results  are only  shown for the
best-fit  model  from the  WMAP3 MCMC, though  the  overall trends are
qualitatively  the same for all other  models, including those for the
WMAP1 cosmology.  At bright magnitude limits, $\langle N \rangle_M$ is
close to a  pure power-law.  At fainter magnitude  limits it starts to
develop a low-mass shoulder,  which, at the faintest magnitude limits,
evolves into a separate  peak.  The upper  middle and upper right-hand
panels plot the  corresponding  occupation statistics of   central and
satellite  galaxies, respectively (see \S\ref{sec:censat}).  The lower
panels plot $\langle N \rangle_M$, plus the contributions from central
(dotted  lines)  and satellite   (dashed  lines) galaxies,   for three
magnitude limits,   as indicated.   Clearly, the  functional  forms of
these  occupation   numbers  are  in  qualitative   agreement with the
functional forms  discussed  above:  $\langle  N_{\rm  cen} \rangle_M$
transits from zero at  low $M$ to  unity at large $M$,  while $\langle
N_{\rm sat} \rangle_M$ is well  approximated  by a power-law at  large
$M$  with an  (exponential) truncation at   low $M$.  However, a  more
detailed  comparison shows  that the   $\langle  N \rangle_M$ of  TNWW
reveal more pronounced shoulders at  $\langle N \rangle_M  = 1$ with a
sharper  truncation at low  $M$. The  sharpness  of this zero-to-unity
transition is a measure for the scatter  in the conditional probability
function $P(M \vert L_{\rm cen})$, which is the topic of discussion in
\S\ref{sec:scatter}

For the  faintest   magnitude  limits   considered here   $\langle   N
\rangle_M$  reveals a small `bump' at  low $M$.  Although the presence
of this  bump is only marginally significant,  in that  there are also
models within the 68 percent confidence levels of the MCMC that do not
reveal such a bump, a similar feature was obtained by Magliocchetti \&
Porciani (2003) from an HOD analysis of the 2dFGRS. It thus seems that
the data has a weak preference for such  a feature, which implies that
a significant fraction of the low mass  haloes host a satellite galaxy
that is    virtually   equally   bright  as the       central  galaxy.
Interestingly, SPH simulations  and semi-analytical models for  galaxy
formation reveal  similar `bumps' in  the  halo occupation statistics,
especially for blue galaxies  (Sheth \& Diaferio 2001; Scranton  2003;
Zheng \etal 2005).

Finally,  in  Fig.~\ref{fig:honfit}  we  compare the  halo  occupation
numbers  predicted from  our CLF  MCMC  with those  obtained from  our
2dFGRS  group catalogue  (cf.   Table~A1).  Overall  the agreement  is
satisfactory,  especially for  brighter magnitude  limits.   The group
catalogue predicts a flattening of $\langle N \rangle_M$ at $\langle N
\rangle_M  \simeq  1$,  in  disagreement  with our  CLF  predictions.  
However,  this is  an artefact  of the  group finder:  one  can detect
``groups''  with $N=1$,  but  not  those with  $N=0$  (cf. Yang  \etal
2005b).

\subsection{Power-law slopes}
\label{sec:slopes}

\begin{figure}
\centerline{\psfig{figure=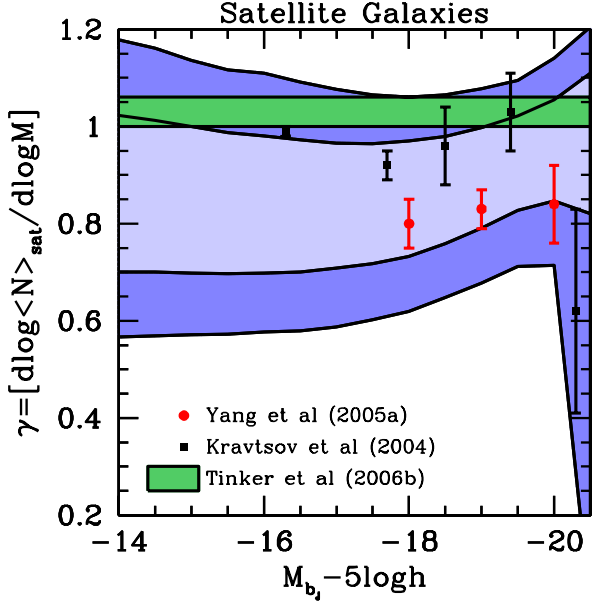,width=0.95\hssize}}
\caption{The slope $\gamma \equiv \rmd\log\langle N_{\rm sat}
  \rangle/\rmd\log M$, measured at $\langle N_{\rm sat} \rangle=3$, as
  function of the magnitude  limit. Overplotted for comparison are the
  results  obtained from  the  2dFGRS group  catalogue  of Yang  \etal
  (2005a), the  results of  Tinker \etal (2006b)  obtained from  a HOD
  analysis of  the 2dFGRS,  and the results  obtained for  dark matter
  subhaloes by Kravtsov \etal (2006b).}
\label{fig:slope}
\end{figure}

To make our predictions regarding the occupation numbers somewhat more
quantitative, we use  our WMAP3 MCMC to compute  the slope of $\langle
N_{\rm sat} \rangle_M$  at the  high-mass end.    Fig.~\ref{fig:slope}
plots the  68 and 95 percent  confidence limits on $\gamma \equiv {\rm
  d}\log  \langle N_{\rm sat} \rangle_M  /  {\rm d}\log M$ measured at
$\langle N_{\rm  sat} \rangle_M  =   3$ as function of  the  magnitude
limit.   This  shows that  there   is a fairly   large  uncertainty on
$\gamma$, especially  for   faint magnitude limits.   In  addition, it
shows that the value of $\gamma$ does not depend strongly on the value
of  the magnitude limit  used.  The  sudden  dramatic increase of  the
confidence limits  at $M_{b_J}-5 \log h  = -20$ owes  to the fact that
for some  of the CLF models  in the MCMC  even the most massive haloes
considered   ($M = 10^{16}    h^{-1}  \Msun$) have  fewer than   three
satellites with $M_{b_J} - 5 \log h \leq -20$, so that $\gamma$ is not
defined.

The  (red) solid  dots correspond  to  the results  obtained from  our
2dFGRS group catalogue,  also measured at $N_{\rm sat}=3$,  and are in
good agreement  with the CLF  constraints. The (green)  horizontal bar
indicates the  constraints on  $\gamma$ obtained by  TNWW from  an HOD
analysis  of  the  2dFGRS.   Under  the assumption  that  $\gamma$  is
independent of the  luminosity limit they obtained $\gamma  = 1.03 \pm
0.03$ (68\% CL). This value  is consistent with our CLF predictions at
the  $1$ to  $2\sigma$ level,  but significantly  higher than  what we
obtained from  our 2dFGRS  group catalogue.  The  same applies  to the
(black)  solid squares, which  indicate the  slopes of  the occupation
statistics of  CDM sub-haloes.  These  have been obtained  by Kravtsov
\etal (2004)  for five different number  densities of CDM  haloes in a
large numerical simulation.  We  have converted these number densities
to a $b_J$ band magnitude  limit, using the 2dFGRS luminosity function
of Madgwick \etal (2002).

Finally, we  emphasize that these  comparisons have to  be interpreted
with some care. After all,  our $\langle N_{\rm sat}\rangle_M$ are not
pure  power-laws,  neither  for  the  CLF  predictions,  nor  for  the
occupation  statistics  obtained  from  the 2dFGRS  group  catalogue.  
Consequently,  the results  depend somewhat  on the  value  of $N_{\rm
  sat}$ at which the slope is measured.

\subsection{Scatter in the relation between $L_{\rm cen}$ and $M$}
\label{sec:scatter}

As mentioned above, the occupation  statistics of TNWW seem to predict
significantly  sharper zero-to-unity  transitions for  $\langle N_{\rm
  cen} \rangle_M$,  which implies  significantly less scatter  in $P(M
\vert  L_{\rm  cen})$.   The  width of  this  conditional  probability
distribution is interesting from  the perspective of galaxy formation,
as it  contains information regarding  the amount of  stochasticity in
galaxy formation.  It is also important for a proper interpretation of
galaxy-galaxy lensing measurements (Mandelbaum  \etal 2005) and of the
kinematics of satellite galaxies (van den Bosch \etal 2004).

We can use the CLF to compute the variance in $\log  M$ of haloes that
host a central galaxy of luminosity $L_{\rm cen}$. This is given by
\begin{equation}
\label{Mcensig}
\sigma^2[\log M] = {\calM_2 \over \calM_0} - 
   \left( {\calM_1 \over \calM_0} \right)^2
\end{equation}
with
\begin{equation}
\label{Mmom}
\calM_k = \int_{0}^{\infty} P(L_{\rm cen} \vert M) \, [\log M]^k 
\, n(M) \, \rmd M
\end{equation}
and     with     $P(L_{\rm     cen}     \vert     M)$     given     by
equation~(\ref{probLbright}).  Here we have used that
\begin{equation}
\label{probLc}
P(M \vert L_{\rm cen}) \rmd M = {P(L_{\rm cen} \vert M) n(M) \over 
\int_{0}^{\infty} P(L_{\rm cen} \vert M) n(M) \rmd M} \rmd M\,,
\end{equation}
which follows from Bayes' theorem.

\begin{figure}
\centerline{\psfig{figure=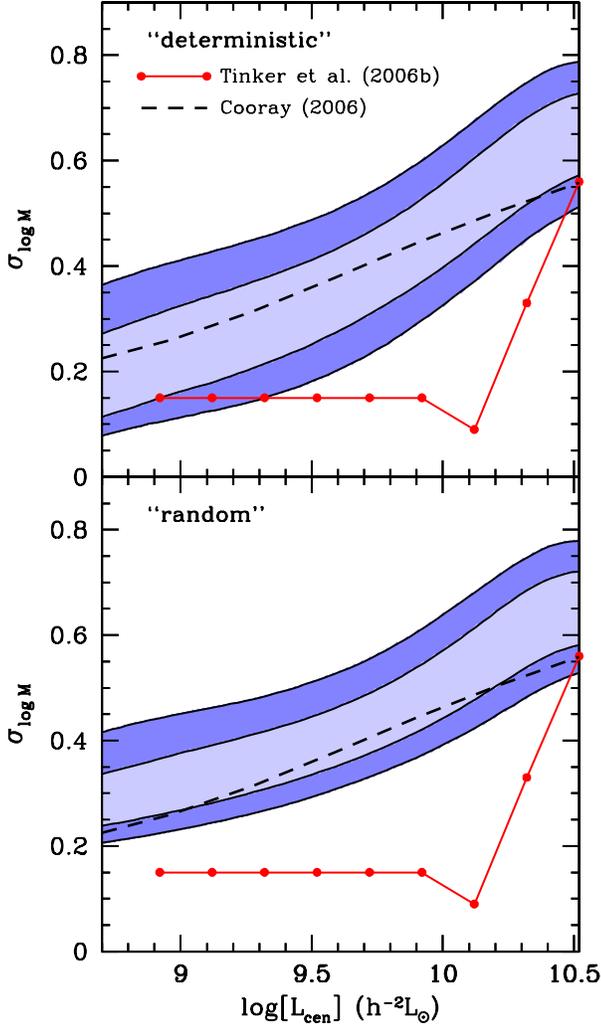,width=0.95\hssize}}
\caption{The contours  show the  68\% and  95\% confidence
  limits on the standard deviation  in $\log M$ as function of $L_{\rm
    cen}$, as  obtained from our  WMAP3 MCMC. This  quantity expresses
  the  width of  the conditional  probability distribution  $P(M \vert
  L_{\rm  cen})$. Upper  and lower  panels  show the  results for  the
  ``deterministic''   and  ``random''   methods,   respectively.   For
  comparison,  we  also show  the  results  obtained  by Tinker  \etal
  (2006b), from an  HOD analysis of the 2dFGRS  (red, solid dots), and
  by Cooray (2006)  from a CLF analysis (dashed  line). Note that both
  CLF studies predict a much broader $P(M \vert L_{\rm cen})$ than the
  HOD study of Tinker \etal.}
\label{fig:sigcen}
\end{figure}

Fig.~\ref{fig:sigcen} plots the 68 and 95 percent confidence limits on
$\sigma_{\log M}(L_{\rm cen})$ obtained  from our WMAP3 MCMC using the
``deterministic'' (upper panel) and ``random'' (lower panel) $P(L_{\rm
  cen} \vert  M)$.  The  CLF model predicts  a pronounced  increase of
$\sigma_{\log M}$  with increasing $L_{\rm cen}$.   This is consistent
with Fig.~\ref{fig:Nstat}, which  clearly shows that the zero-to-unity
transition  of $\langle N_{\rm  cen} \rangle$  becomes less  sharp for
brighter  magnitude limits.  As  expected, the  scatter in  $P(M \vert
L_{\rm  cen})$  is higher  in  the  ``random''  case compared  to  the
``deterministic'' case, especially at the faint end.

The  dashed  lines  in  Fig.~\ref{fig:sigcen}  indicate  $\sigma_{\log
  M}(L_{\rm cen})$  for the $B$-band  CLF from Cooray  (2006) computed
using~(\ref{Mcensig}).  Although Cooray assumes a different functional
form for the CLF, and uses a very different technique to constrain the
CLF, the agreement with our results is remarkably good.  

The (red)  solid dots show  the results obtained  by TNWW from  an HOD
analysis of the 2dFGRS.  For $L_{\rm cen} < 10^{10} h^{-2} \Lsun$ TNWW
have  {\it  assumed} that  $\sigma_{\log  M}  =  0.15$.  For  brighter
centrals, however,  they let  $\sigma_{\log M}$ be  a free  parameter. 
Their best-fit values show a strong increase of $\sigma_{\log M}$ with
increasing $L_{\rm  cen}$. Although in qualitative  agreement with our
results and those of Cooray (2006), their values for $\sigma_{\log M}$
are much smaller. Unfortunately, TNWW do not give any uncertainties on
their best-fit $\sigma_{\log M}$, so that it is difficult to judge the
significance of this difference.

Given  the relevance of  the amount  of scatter  in $P(M  \vert L_{\rm
  cen})$ for,  for example,  weak lensing, it  is important to  try to
obtain  more direct constraints  on $\sigma_{\log  M}$. In  More \etal
(2006,  in preparation),  we  use satellite  kinematics  to show  that
$\sigma_{\log M}(L_{\rm cen}) > 0.2$  for $L_{\rm cen} > 3 \times 10^9
h^{-2} \Lsun$,  which clearly rules  out the relatively  small scatter
obtained (and  assumed) by TNWW.  It  is unclear at  present why these
authors  obtain  a $\sigma_{\log  M}(L_{\rm  cen})$  that  is so  much
smaller.  For example, as shown in Zehavi \etal (2005), one can change
the sharpness of the  zero-to-unity transition of $\langle N_{\rm cen}
\rangle_M$ and leave the fit to the galaxy-galaxy correlation function
largely intact.

\section{Pairwise Velocity Dispersions}
\label{sec:pvd}

The peculiar  velocities of galaxies are determined   by the action of
the gravitational field, and  are  therefore directly related to   the
matter distribution  in the Universe.   Consequently, the amplitude of
galaxy peculiar  velocities can  yield useful,  additional information
regarding   the universal  relation    between  light and   mass.  One
statistic that is particularly useful  in this respect is the pairwise
velocity dispersion (PVD), $\sigma_{12}(r)$, which is a measure of the
relative peculiar  velocity of a  pair  of galaxies as a  function  of
their separation  $r$.   The PVDs can  be  obtained from the   data as
described in \S\ref{sec:methodpvd} below.

In Yang  \etal (2004)  we used detailed  mock galaxy  redshift surveys
(hereafter MGRSs,  see Appendix~B) constructed using our  WMAP1 CLF in
order to  investigate what  these CLF models  predict for the  PVDs of
2dFGRS  galaxies.  A  comparison  with the  results  of Hawkins  \etal
(2003) revealed  that  our  MGRS  {\it  based  on  the  best-fit}  CLF
dramatically over-predicts the PVDs at  scales of $\sim 1 h^{-1} \Mpc$
by $\sim  350 \kms$. Since the  PVD is extremely sensitive  to the few
richest systems  in the sample (i.e.,  Mo, Jing \&  B\"orner 1993) one
can  lower the  PVDs by  lowering  the occupation  numbers of  massive
haloes.  Within the  uncertainties of the CLF parameters  we were able
to find a model that  could reproduce the observed PVDs. However, that
model predicts  an average mass-to-light  ratio for clusters  of $\sim
1000 h\Msun/\Lsun$, which is much larger (by more than $7\sigma$) than
the  average  mass-to-light  ratio  obtained from  other,  independent
measurements (see \S\ref{sec:ml}).

Rather than  lowering the average number  of galaxies per cluster, one
can also  lower the PVDs by reducing  the abundance of massive haloes.
This, however, implies a change of cosmology.  In Yang \etal (2004) we
showed  that a  flat  $\Lambda$CDM  cosmology with $\Omega_m=0.3$  and
$\sigma_8=0.75$  could accurately reproduce  the  observed PVDs with a
realistic mass-to-light ratio  for clusters.   In  fact, we used  this
result    to argue against  the  WMAP1  cosmology  and in   favor of a
cosmology  with  a reduced $\sigma_8$   (see also van  den Bosch \etal
2003b, 2005a and Yang \etal 2005a).

The    PVDs of  Hawkins  \etal  (2003)    were obtained  from  a large
flux-limited  sample.   Although    these already  provide   important
constraints  on  the  mass-to-light ratios of  clusters   (for a given
cosmology),  one can obtain even tighter  constraints on the universal
relation between   light and mass by  measuring   the PVDs {\it   as a
  function  of galaxy luminosity}.   Jing \& B\"orner (2004, hereafter
JB04)   were the first  to  present  a  PVD  analysis  for galaxies in
different luminosity intervals.  Using the 2dFGRS, they found that the
PVD at a scale of $k = 1 h \Mpc^{-1}$ has a minimum of $\sim 425 \kms$
for galaxies with $M_{b_J}  - 5 \log h  \sim -20.5$. Fainter galaxies,
with magnitudes  in the range  $[-17,-19]$,  were found to  have  much
higher pairwise velocity dispersions of $\sim 700 \kms$, almost as high
as those  of the brightest  galaxies  in  the 2dFGRS (see  red,  solid
triangles   in    Fig.~\ref{fig:sig12k}).    This indicates    that  a
significant  fraction of   the   fainter galaxies  must   be satellite
galaxies in massive haloes.

A qualitatively similar  result has been obtained  by  Li \etal (2006;
hereafter L06) from an analysis of the SDSS.   In their case, however,
$\sigma_{12}(L)$ measured at  $k = 1 h   \Mpc^{-1}$ reveals a  smaller
dynamic range; the minimum occurs  at $\sim 500  \kms$ while the  PVDs
for the fainter galaxies are $\lta 600  \kms$ (see blue, solid squares
in Fig.~\ref{fig:sig12k}).

Using  one of  the CLF  models presented  in Yang  \etal  (2003), JB04
constructed a mock 2dFGRS which  they analyzed in exactly the same way
as the 2dFGRS  data.  Contrary to the data, the  model PVDs were found
to increases monotonically from $\sim 400 \kms$ at $M_{b_J} - 5 \log h
= -17$ to $\sim 750 \kms$ at $M_{b_J} - 5 \log h = -21$. The CLF model
thus severely underpredicts  the PVDs of faint galaxies,  and does not
reveal the  pronounced minimum near $M_{b_J}  - 5 \log h  \sim -20.5$. 
However,  there  is  a  considerable  amount of  freedom  in  the  CLF
parameters.   For example,  as is  evident  from Figs.~\ref{fig:histo}
and~\ref{fig:ml}, within  the 95 percent confidence limits  there is a
wide  range of  cluster mass-to-light  ratios and  satellite fractions
that  can  fit  both  $\Phi(L)$  and $r_0(L)$.   Since  the  PVDs  are
extremely sensitive  to these quantities (e.g., Mo  \etal 1993; Slosar
\etal 2006;  Tinker 2006c),  it is crucial  that one takes  this model
freedom into account  when comparing model and data.   Here we will do
so, analysing the luminosity dependence  of the PVDs for our new WMAP3
CLF.
\begin{table*}
\caption{Pairwise Velocity Dispersions}
\begin{tabular}{cccccc}
\hline
  Magn. Limits    &   Median Magn.    & $z_{min}$ & $z_{\rm max}$ & $N$ & $\sigma_{12}(k=1h\Mpc^{-1})$ \\
$M_{b_J}-5\log h$ & $M_{b_J}-5\log h$ &           &               &     & $\kms$  \\
\hline\hline
$\langle -17.5,-16.5]$ & $-16.99$ & $0.01$ & $0.05$ &  $4892$ & $507 \pm 122$ \\
$\langle -18.0,-17.0]$ & $-17.48$ & $0.01$ & $0.06$ &  $8144$ & $532 \pm 82$  \\
$\langle -18.5,-17.5]$ & $-17.99$ & $0.01$ & $0.07$ & $12525$ & $592 \pm 38$  \\
$\langle -19.0,-18.0]$ & $-18.49$ & $0.01$ & $0.09$ & $24334$ & $574 \pm 52$  \\
$\langle -19.5,-18.5]$ & $-18.96$ & $0.02$ & $0.11$ & $35461$ & $527 \pm 43$  \\
$\langle -20.0,-19.0]$ & $-19.43$ & $0.02$ & $0.13$ & $41438$ & $470 \pm 48$  \\
$\langle -20.5,-19.5]$ & $-19.90$ & $0.02$ & $0.16$ & $43600$ & $451 \pm 44$  \\
$\langle -21.0,-20.0]$ & $-20.36$ & $0.04$ & $0.20$ & $36383$ & $413 \pm 28$  \\
$\langle -21.5,-20.5]$ & $-20.79$ & $0.05$ & $0.20$ & $12853$ & $694 \pm 134$ \\
$\langle -22.0,-21.0]$ & $-21.24$ & $0.06$ & $0.20$ &  $2840$ & $993 \pm 289$ \\
\hline
\end{tabular}
\medskip

\begin{minipage}{\hdsize}
  Column~(1) specifies  the absolute  magnitude limit for  each volume
  limited sample, while the median  magnitude is listed in column~(2). 
  The  minimum and  maximum redshifts  of  each sample  are listed  in
  columns~(3) and  (4), respectively, and  the total number  of 2dFGRS
  galaxies in  each sample  is listed in  column~(5).  Note  that only
  galaxies with a redshift completeness greater than 0.7 are selected.
  Finally, column (6) lists the PVDs  in the 2dFGRS measured at $k=1 h
  \Mpc^{-1}$,  plus  the (cosmic  variance)  error  determined from  8
  MGRSs.
\end{minipage}
\end{table*}

\subsection{The Luminosity Dependence of the PVDs in the 2dFGRS}
\label{sec:methodpvd}

We start by  performing our own analysis of  the luminosity dependence
of the PVDs in the 2dFGRS. First we select those galaxies in the final
release  of the 2dFGRS  that are  located in  the North  Galactic Pole
(NGP)  and South  Galactic Pole  (SGP) survey  strips with  a redshift
quality parameter $Q \geq 3$, with $0.01 \leq z \leq 0.20$, and with a
redshift  completeness $\ge  0.7$.  These are  used  to construct  ten
volume-limited samples (adopting  $b_J=19.3$ as the apparent magnitude
limit of the 2dFGRS) whose magnitude and redshift limits are indicated
in Table~3.

Let $r_p$ and $\pi$ be the pair separations perpendicular and parallel
to  the line-of-sight,  respectively. For  each of  our volume-limited
samples   we  compute  the   two-point  correlation   function  (2PCF)
$\xi(r_p,\pi)$,  using the  estimator introduced  by Hamilton  (1993). 
Random  samples are  constructed from  our MGRSs  (see  Appendix~B) by
randomizing  the  coordinates  of  all  mock  galaxies.  We  use  this
two-dimensional 2PCF to compute the PVD from the galaxy power spectrum
in  redshift space, $P^{(s)}(k,\mu)$,  which is  related to  the power
spectrum in real space, $P(k)$ according to
\begin{equation}
\label{Pstheory}
P^{(s)}(k,\mu) = P(k) \, (1 + \beta \mu^2)^2 \, D[k \mu \sigma_{12}(k)]
\end{equation}
(Peacock \&  Dodds 1994; Cole, Fisher  \& Weinberg 1995)  with $k$ the
wavenumber and  $\mu$ the cosine  of the angle between  the wavevector
and the line-of-sight.  The factor  $(1 + \beta \mu^2)^2$ accounts for
the   compression   due   to   infall,   with   $\beta$   the   linear
redshift-distortion parameter, while  $D[k \mu \sigma_{12}(k)]$ is the
damping  function that  accounts  for the  random  motion of  galaxies
within dark  matter haloes.   We follow JB04  and L06 and  assume that
this damping function has a Lorentz form
\begin{equation}
\label{lorentz}
D[k \mu  \sigma_{12}(k)] = \left[ 1 + {1 \over 2} k^2 \mu^2
  \sigma_{12}^2(k) \right]^{-1}\,,
\end{equation}
and compute the redshift space  power spectrum for each volume limited
sample in Table~3 by Fourier transforming the corresponding 2PCF:
\begin{eqnarray}
\label{Psdata}
P^{(s)}(k,\mu) & = & 2 \pi \int \rmd\pi \int \rmd r_p \, r_p \,
\xi(r_p,\pi) \, \cos(k_{\pi}\pi) \nonumber\\
 & & J_0(k_p r_p) \, W(r_p,\pi)
\end{eqnarray}
(JB04)\footnote{Don't confuse the $\pi$'s  in this equation: The first
  $\pi$ is  the usual $3.14159...$,  while the other  $\pi$'s indicate
  the  separation  along  the  line-of-sight}  .  Here  $J_0$  is  the
zeroth-order   Bessel  function,   $k_{\pi}$  and   $k_{p}$   are  the
wavenumbers perpendicular and parallel to the line-of-sight, and
\begin{equation}
W(r_p,\pi) = \exp\left(-{r_p^2 + \pi^2 \over 2 S^2} \right)
\end{equation}
is a  Gaussian smoothing function  (with smoothing scale  $S=20 h^{-1}
\Mpc$)  which  is used  to  suppress  the  impact of  fluctuations  in
$\xi(r_p,\pi)$ at large separations (see JB04 for details). We compute
$\xi(r_p,\pi)$ in  equal logarithmic bins  of $r_p$ ($\Delta  {\rm ln}
r_p  = 0.23$) and  in equal  linear bins  of $\pi$  ($\Delta \pi  = 1.0
h^{-1} \Mpc$).  The $\pi$-integral in (\ref{Psdata}) is performed over
the interval $-50 h^{-1} \Mpc \leq  \pi \leq 50 h^{-1} \Mpc$, while we
integrate $r_p$ from $0.1 h^{-1} \Mpc$ to $50 h^{-1} \Mpc$.

\begin{figure*}
\centerline{\psfig{figure=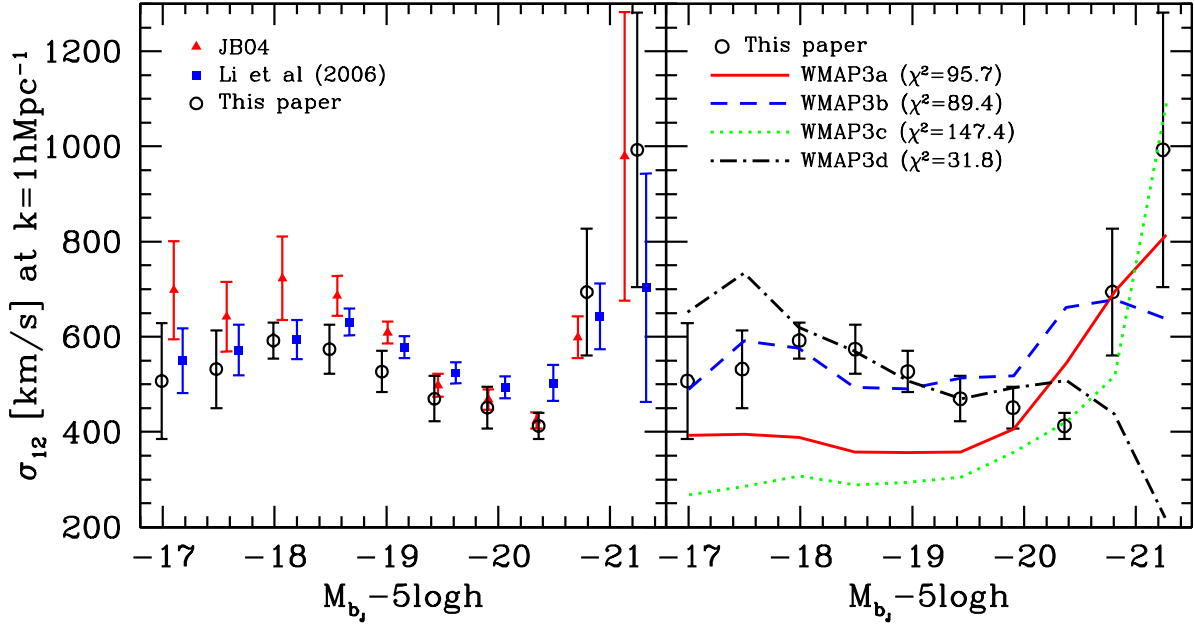,width=0.9\hdsize}}
\caption{The PVD    measured at $k=1  h  \Mpc^{-1}$  using the Fourier
  analysis described in   \S\ref{sec:methodpvd}  as  function   of the
  median magnitude of the volume limited sample used.  Open circles in
  both panels   indicate the results  obtained in  this paper from the
  2dFGRS  (see  Table~3), with  the  errorbars  indicating  the cosmic
  variance as  obtained from  8  MGRSs.   In the left-hand  panel,  we
  compare these to the results obtained by JB04 (red, solid triangles)
  and L06 (blue, solid  squares) from similar  analyses of the  2dFGRS
  and SDSS, respectively.    We  have  converted the SDSS     $r$-band
  magnitudes to $b_J$  band magnitudes assuming  $b_J - r =  0.9$ (see
  L06).   In  the right-hand panel,  we compare  our 2dFGRS results to
  those  obtained from four different  WMAP3  CLF models, as indicated
  (cf. Table~2 and  Fig.~\ref{fig:models}).  For completeness, we have
  indicated  the formal $\chi^2$ value for  each  of these models. See
  text for detailed discussion.}
\label{fig:sig12k}
\end{figure*}

Finally we determine the real-space power spectrum $P(k)$ and the PVDs
$\sigma_{12}(k)$  by  modeling  the  measured $P^{(s)}(k,\mu)$  using
eq.(\ref{Pstheory})  with $\beta = 0.45$.   Detailed tests in JB04 and
L06   have shown that  keeping  $\beta$   fixed  at this value  yields
reliable results.    The  best-fit    values  for $\sigma_{12}(k=1   h
\Mpc^{-1})$ thus   obtained are listed  in  Table~3  and are  shown in
Fig.~\ref{fig:sig12k} as  black,   open  circles.  The   errorbars are
obtained  from 8   mock  redshift surveys (see  \S\ref{sec:comparison}
below) and indicate the expected scatter due to cosmic variance.

Comparison of these results with  those of JB04  and L06, reveals good
mutual agreement at  $M_{b_J}   - 5\log  h   \gta -19$.   For  fainter
galaxies,  however, our analysis yields PVDs  that are $\sim 150 \kms$
lower than those of JB04, with the results  of L06, which are based on
the SDSS, roughly in between.  Since our analysis is identical to that
of JB04, these   differences reflect the slightly different  selection
criteria.  Whereas JB04  used flux-limited samples  with $0.02 \leq  z
\leq  0.25$, we use volume-limited  samples with the restrictions that
$0.01 \leq z \leq 0.20$.  Another  potential source of this difference
is the  relative sensitivity to the exact  scale at which the PVDs are
measured.  As can be seen from Fig.~7  in JB04, their $\sigma_{12}(k)$
for galaxies   with $-18.5 < M_{b_J}   - 5\log h  < -17.5$   reveals a
pronounced,  sharp peak  of $\sim  725   \kms$ at $k=1 h   \Mpc^{-1}$.
However, at slightly higher or lower $k$, the PVDs are $\sim 550 \kms$
in much better agreement with our results and those of L06.

\subsection{Comparison with CLF Models}
\label{sec:comparison}

In order  to predict  PVDs from our  CLF, we construct  detailed MGRSs
using  the  CLF and  cosmological  $N$-body  simulations (dark  matter
only).  These  MGRS are constructed  to be directly comparable  to the
2dFGRS, as described  in detail in Appendix~B. We  analyze these MGRSs
using exactly  the same  procedure (described above)  as used  for the
2dFGRS data, so that the model-data comparison is as fair as possible.

The solid (red) line  in the right-hand panel of Fig.~\ref{fig:sig12k}
indicates  the PVDs  obtained from   the   MGRS constructed from   the
best-fit  CLF (called WMAP3a in Table~2).   For $M_{b_J}-5 \log h \lta
-19.5$ this model predicts PVDs  that are in reasonable agreement with
the data.  For fainter samples, however,  the PVDs are clearly too low
compared to the  2dFGRS.  This is  in qualitative agreement with JB04,
even though our analysis is for the WMAP3 cosmology, while that of JB04
was for a WMAP1 cosmology.

\begin{figure*}
\centerline{\psfig{figure=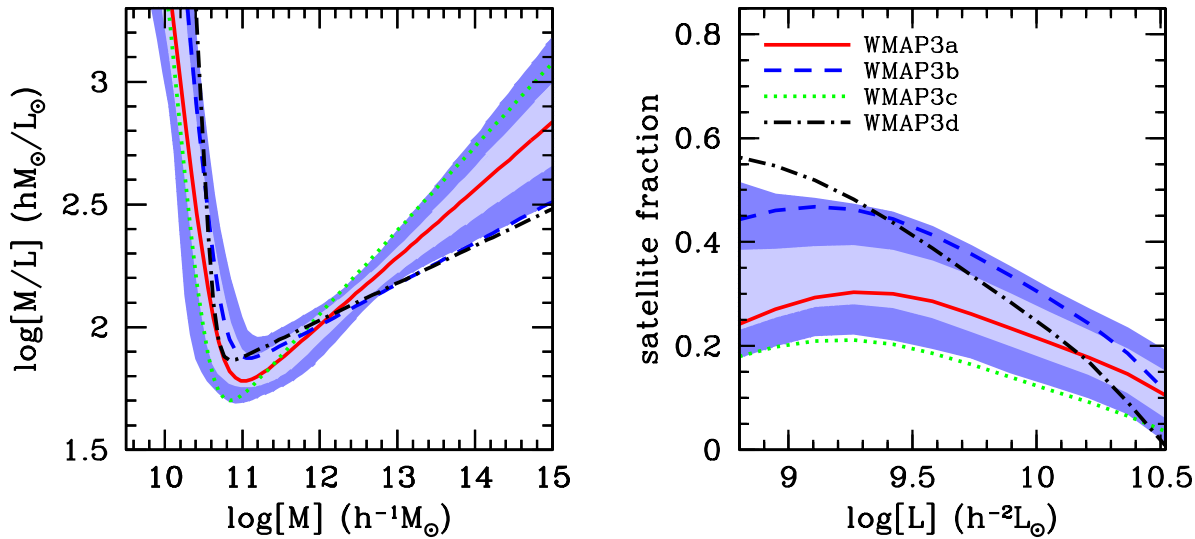,width=0.9\hdsize}}
\caption{Some predictions of four models discussed  in the text (lines)
  overplotted  on  the  68\%  and  95\%  confidence  limits  from  the
  marginalized distributions  of the WMAP3 MCMC.   The left-hand panel
  shows  the average  mass-to-light ratio  as function  of halo  mass. 
  Contrary to  Fig.~\ref{fig:ml}, here we plot  $\langle M/L \rangle$,
  not $\rangle  M/L_{18} \langle$, with  $\langle L \rangle$  given by
  equation~(\ref{ltot}).   The right-hand  panel  shows the  satellite
  fraction as function of galaxy luminosity.  Model WMAP3a corresponds
  to the  best-fit model in the  MCMC, while models  WMAP3b and WMAP3c
  roughly  outline  the extrema  of  $\langle  M/L  \rangle_M$ and  of
  $f_{\rm sat}(L)$.  Finally,  model WMAP3d, whose mass-to-light ratio
  is  almost identical  to that  of  WMAP3b, has  an extremely  strong
  gradient  in $f_{\rm sat}(L)$.   As shown  in Fig.~\ref{fig:sig12k},
  this is  the model that best  fits the luminosity  dependence of the
  PVDs.}
\label{fig:models}
\end{figure*}

In  order to  probe  the  uncertainties on  $\sigma_{12}$  due to  the
uncertainties  on the CLF  parameters, ideally  one would  construct a
MGRS  for each of  the 2000  models in  our MCMC.   Unfortunately, the
construction   of   MGRSs    and   their   subsequent   analysis,   is
computationally too expensive, rendering this unpractical. Instead, we
proceed as follows. Since the  mass-to-light ratio of clusters and the
satellite  fractions are  the  two model  aspects  that most  strongly
impact on $\sigma_{12}$, we have searched the MCMC for two models that
more  or less  bracket the  95 percent  confidence limits  of  our CLF
model.  The parameters of these  models, called WMAP3b and WMAP3c, are
listed in Table~2, while Fig.~\ref{fig:models} shows the mass-to-light
ratios and satellite fractions of these models.

The dashed  (blue) and  dotted (green) lines  in Fig.~\ref{fig:sig12k}
show the PVDs of models WMAP3b and WMAP3c, respectively.  Model WMAP3b
predicts  significantly higher satellite  fractions and  lower cluster
mass-to-light ratios than  the best-fit model (WMAP3a).  Consequently,
the PVDs  for faint  galaxies are much  larger, bringing them  in much
better agreement with the PVDs  obtained from the 2dFGRS.  In order to
quantify  the  comparison between  different  models,  we compute  the
formal $\chi^2$  using the  `cosmic variance' errorbars  obtained from
our 8 mock redshift surveys. This yields $\chi^2$ values of $95.7$ and
$89.4$ for  models WMAP3a and WMAP3b, respectively.  Despite the clear
improvement at the  faint end, the goodness-of-fit of  model WMAP3b is
only  marginally  better than  for  model  WMAP3a.   This owes  almost
entirely to the  fact that model WMAP3b severely  overpredicts the PVD
for  galaxies with $M_{b_J}-5\log  h \simeq  -20.4$: this  single data
point  contributes $80.9$ to  the total  $\chi^2$! 

As expected,  model WMAP3c predicts PVDs  that are even  lower than in
the case of model WMAP3a, in clear disagreement with the data ($\chi^2
=   147.4$).   It  does,   however,  accurately   match  the   PVD  at
$M_{b_J}-5\log h  \simeq -20.4$.  This  suggests that perhaps  a model
with a high $f_{\rm sat}$ at the faint end, and a low $f_{\rm sat}$ at
the high end,  could fit the PVDs at  all luminosities.  Model WMAP3d,
which  we   extracted  from  our  MCMC,  meets   these  criteria  (see
Fig.~\ref{fig:models}), and indeed yields  PVDs that are in reasonable
agreement  with the  data  ($\chi^2 =  31.8$).   It does  dramatically
underpredict the PVDs  at the bright end, but  since the corresponding
(cosmic  variance) errors  are  huge, the  contribution  to the  total
$\chi^2$ is only modest. 

Thus we conclude  that, within the WMAP3 cosmology,  one can find halo
occupation models  that can provide a reasonable,  simultaneous fit to
the  luminosity   dependence  of  the  clustering   strength  and  the
luminosity dependence of  the pairwise velocity dispersions.  However,
this {\it  does} come at a  price.  The best-fit model  (WMAP3d) is an
extreme model within  the MCMC; this is evident  from both Table~2 and
Fig.~\ref{fig:models},  which   show  that  model   WMAP3d  has  model
parameters, mass-to-light  ratios and satellite  fractions that differ
substantially from the best-fit  model.  Furthermore, this model still
does not fit the PVDs  completely satisfactory.  In particular it does
not reveal  a pronounced minimum in $\sigma_{12}(L)$,  as observed. In
fact, we have tested a number  of additional models from our MCMC with
similar $f_{\rm sat}(L)$ as model  WMAP3d, but none fair any better in
this respect. We have also  tested models with a `random', rather than
a `deterministic' sampling of $L_{\rm  cen}$, but this does not have a
significant impact on the PVDs. Although we only tested a hand-full of
models (selected in a strongly biased way), we therefore conclude that
the detailed shape of the  luminosity dependence of the PVDs remains a
challenge  for the  halo occupation  models.  In  the next  section we
discuss possible implications of these findings.

\section{Conclusions}
\label{sec:concl}

Using the  conditional luminosity  function (CLF) formalism,  and data
from the  2dFGRS, we have  constrained the universal  relation between
light and mass. Using a Monte-Carlo Markov Chain we probe the complete
parameter space  of our models,  and provide confidence limits  on all
derived quantities. With  respect to our previous CLF  studies we have
made the following changes and improvements:
\begin{itemize}

\item We have  adopted a flat  $\Lambda$CDM cosmology  with parameters
  advocated by the 3 year data release from the WMAP mission.

\item We have modeled the 2dFGRS data on its light cone.

\item We have taken the scale dependence of the halo bias into account.

\item We no longer impose that the mass-to-light ratios of haloes with
$M \geq 10^{14} h^{-1} \Msun$ are constant.

\end{itemize}
The change in cosmology  (lower $\Omega_m$, lower $\sigma_8$ and lower
spectral index) causes a reduction in the mass-to-light ratios of dark
matter haloes ranging  from $\sim 25$ to $\sim  45$ percent, depending
on the mass  scale. As anticipated, this removes  an important problem
with  previous CLF  and HOD  models which  had a  tendency  to predict
mass-to-light ratios  for clusters that  were too high (van  den Bosch
\etal  2003b, 2005b;  Yang  \etal  2004; Tinker  \etal  2005; Vale  \&
Ostriker 2006).

Taking account of the light-cone  modeling and the scale dependence of
the halo  bias only has  a mild impact  on our results,  improving the
accuracy of  our models  by $\sim 5$  to $10$ percent.   We emphasize,
though, that the impact of these effects can be much larger when using
data out to higher redshift,  or when using clustering data on smaller
scales, compared to what we have used here.

We have  compared various  predictions of our  CLF model  with results
obtained from our 2dFGRS group catalogue. We found excellent agreement
for  the average  mass-to-light  ratios, the  luminosities of  central
galaxies as function of halo mass, the faint-end slope of the CLF, the
occupation numbers in various  luminosity bins, and the luminosity-gap
statistic.  The fact that these two completely different methods yield
results  in  such good  agreement,  and for  such  a  wide variety  of
statistics, is a major success for  both the CLF formalism and for the
halo-based group  finder of  Yang \etal (2005a).   The CLF  model also
predicts  that  the   satellite  fraction  decreases  with  increasing
luminosity,  in excellent agreement  with the  HOD analyses  of Tinker
\etal  (2006b) and  Cooray (2006),  as  well as  with the  constraints
obtained  by  Mandelbaum \etal  (2006)  from  a galaxy-galaxy  lensing
analysis of the SDSS.

One outstanding  issue regarding the mass-to-light  ratios regards the
actual slope  of $\langle M/L \rangle_M$  at the massive  end ($M \gta
10^{14}  h^{-1}  \Msun$). While  the  group  catalogue  of Yang  \etal
(2005a) yields mass-to-light ratios  that continue to increase roughly
as $\langle M/L \rangle_M \propto  M^{\gamma}$ with $\gamma = 0.33 \pm
0.05$, an alternative group catalogue  by Eke \etal (2004), also based
on the  2dFGRS, predicts  that $\gamma \rightarrow  0$ at  the massive
end. Unfortunately, the clustering data  used to constrain the CLF can
not  discriminate  between  these   different  values  for  $\gamma$.  
Although   recent,  independent   studies  seem   to   favor  somewhat
intermediate  values of $\gamma  \simeq 0.2  \pm 0.08$  (e.g., Popesso
\etal 2005), the  fact that two group catalogues  constructed from the
same  data set yield  such wildly  different results,  accentuates the
need for more thorough investigations.

We also presented  a detailed description of the  link between the CLF
and the more  often used HOD models. In particular,  we have shown how
to compute the full halo-occupation distribution, $P(N \vert M)$, from
the CLF  {\it for  any range in  luminosities}.  In addition,  we have
compared  the shape  of $\langle  N  \rangle_M$ predicted  by our  CLF
models with that typically assumed in HOD models.  Although they agree
qualitatively,  the   HOD  models  typically   adopt  a  zero-to-unity
transition for $\langle  N \rangle_M$ which is much  sharper than what
we  predict  with  our  CLF.   This  implies  that  the  CLF  predicts
probability  distributions  $P(M \vert  L_{\rm  cen})$  that are  much
broader than what  is typically assumed in HOD  models.  The amount of
scatter in  $P(M \vert  L_{\rm cen})$ plays  an important role  in the
interpretation   of  weak  lensing   measurements  and   of  satellite
kinematics. In More  \etal (2006, in preparation) we  present a strict
lower limit on $\sigma_{\log  M}$, obtained from satellite kinematics,
which rules out values for $\sigma_{\log M}$ lower than $\sim 0.2$.

Finally,  we have studied  the luminosity  dependence of  the pairwise
velocity  dispersions, $\sigma_{12}$, of  2dFGRS galaxies.   Using ten
volume  limited samples,  we  obtain that  $\sigma_{12}(L)$ reveals  a
local minimum at $M_{b_J}-5 \log  h \simeq -20.4$, in good qualitative
agreement with  Jing \& B\"orner (2004)  and Li \etal  (2006).  At the
faint end, however, we obtain  PVDs that are $\sim 150\kms$ lower than
those of  JB04. Since we used  exactly the same  analysis technique as
JB04, these differences must reflect the different selection criteria.
Using  detailed  mock  galaxy   redshift  surveys  we  compared  these
$\sigma_{12}(L)$ with  predictions from our CLF  models.  In agreement
with previous studies (e.g., Mo  \etal 1993; Slosar \etal 2006; Tinker
2006c) we find that the  PVDs are extremely sensitive to the satellite
fractions,  $f_{\rm  sat}(L)$,  and  to  the  (cluster)  mass-to-light
ratios. This is good news  since the clustering data used to constrain
the  CLF   leaves  relatively  large   uncertainties  regarding  these
quantities.   Simultaneously  matching  $r_0(L)$  and  $\sigma_{12}(L)$
therefore allows  us to strongly tighten the  constraints on parameter
space. In particular,  for the WMAP3 cosmology used  here we find that
$\sigma_{12}(L)$  requires models  with  relatively low  mass-to-light
ratios  for  clusters ($\langle  M/L  \rangle_{\rm  cl}  \simeq 215  h
\MLsun$) and with  a satellite fraction that decreases  from $\sim 45$
percent at $M_{b_J}-5 \log h=-18.5$ to $\sim 10$ percent at $M_{b_J}-5
\log h=-20.5$.

In terms of the likelihood distributions obtained from our MCMC, these
values are  typically $\gta 2\sigma$ away from  the median, indicating
that our CLF  model is not capable of  accurately fitting $r_0(L)$ and
$\sigma_{12}(L)$ simultaneously. In particular, we were unable to find
a CLF model in our MCMC that could reproduce the pronounced minimum in
$\sigma_{12}(L)$ at  $M_{b_J}-5 \log  h \simeq -20.4$.   This suggests
either (i) that we are dealing  with the wrong cosmology, or (ii) that
the  CLF  parameterization used  here  is  not  sufficiently general.  
Although we  certainly can't rule  out this latter option,  we believe
that option (i) is the more likely cause for this outstanding problem.
This  is motivated by  some of  our previous  results.  In  Yang \etal
(2004) we  used our  CLF formalism  and the  PVDs measured  by Hawkins
\etal (2003)  to argue against the  WMAP1 cosmology and in  favor of a
cosmology with  $\Omega_m \simeq 0.3$ and $\sigma_8  \simeq 0.75$. The
WMAP3 cosmology adopted  here, has $\Omega_m = 0.238$  and $\sigma_8 =
0.744$.  Lowering  $\Omega_m$ and/or $\sigma_8$  reduces the abundance
of  massive   haloes,  which  in  turn  implies   that  lower  cluster
mass-to-light ratios are needed in order to explain the observed PVDs.
The fact that the WMAP3 cosmology studied here requires relatively low
values  for $\langle M/L  \rangle_{\rm cl}$,  while our  WMAP1 studies
required  relatively  high  cluster  mass-to-light  ratios,  therefore
suggests  a cosmology  with  values for  $\Omega_m$ and/or  $\sigma_8$
intermediate  between those of  the WMAP1  and WMAP3  cosmologies.  We
leave it for future studies to see whether indeed such a cosmology can
yield  a   CLF  that  can   simultaneously  match  the   $r_0(L)$  and
$\sigma_{12}(L)$ with  realistic model  parameters.  As a  final note,
however,  we wish  to  emphasize that  the  combined constraints  from
$r_0(L)$ and $\sigma_{12}(L)$ are extremely tight, thus offering great
potential   to  constrain  both   cosmological  parameters   and  halo
occupation statistics.   The CLF  formalism presented here  is ideally
suited for such a task.


\section*{Acknowledgments}

FvdB  acknowledges  useful  discussions  with Andreas  Berlind,  Aaron
Dutton, Andrey  Kravtsov and Jeremy  Tinker.  The simulations  used in
this  paper   have  been  carried  out  on   the  zBox2  supercomputer
(http://www-theorie.physik.unizh.ch/~dpotter/zbox2/)  at the Institute
for Theoretical Physics in Zurich, Switzerland.



\appendix

\section[]{The 2dFGRS Group Catalogue}
\label{sec:AppA}

Throughout this paper, we compare various CLF predictions with results
obtained from  our 2dFGRS group  catalogue.  The construction  of this
catalogue is based on a halo-based group finder, which is described in
Yang \etal (2005a, hereafter YMBJ).  This group finder is optimized to
group  together those  galaxies that  reside in  the same  dark matter
halo, and has been tested in great detail against mock galaxy redshift
surveys (YMBJ; Yang \etal 2005b; Weinmann \etal 2006a,b)

Contrary to most  other studies, we do not  determine the group masses
from  the  velocity dispersion  of  the  group  members.  Instead,  we
estimate the group masses  from the group luminosity $L_{18}$, defined
as  the   total  luminosity  of   all  group  members   brighter  than
$M_{b_J}-5\log h = -18$. Detailed tests have shown that this method is
significantly  more accurate  than  using the  velocity dispersion  of
group members  (see Appendix~B of Weinmann \etal  2006a).  For distant
groups,  in which  not all  galaxies  above this  magnitude limit  are
brighter than  the flux limit of  the survey, we  correct $L_{18}$ for
the missing  members using an empirical self-calibration  based on the
groups that are sufficiently  nearby (see YMBJ for details).  Finally,
to convert  from $L_{18}$ to a  halo mass $M$, we  make the assumption
that there  is a  one-to-one relation between  $L_{18}$ and  $M$.  For
each group we determine the  number density of all groups brighter (in
terms of  $L_{18}$) than the group  in consideration, and  we then use
the halo  mass function for the  WMAP3 cosmology to find  the value of
$M$ for  which the more massive  haloes have the same  number density. 
Note that this  has the disadvantage that the  group mass is cosmology
dependent.   However,  it  can   easily  be  converted  to  any  other
cosmology, using the relation
\begin{equation}
\label{massconvert}
\int_{M}^{\infty} n(M') \rmd M' = \int_{\widetilde{M}}^{\infty} 
\tilde{n}(M') \rmd M' 
\end{equation}
Here $M$ and  $n(M)$ are the mass and  halo mass function in the WMAP3
cosmology,     and   $\widetilde{M}$   and   $\tilde{n}(M)$   are  the
corresponding values in the other cosmology .
\begin{table}
\label{tab:masslight}
\caption{Parameters of Galaxy Groups in 2dFGRS}
\begin{tabular}{crrccc}
\hline
  $\log M$  &  $\log\langle L_{\rm cen}\rangle$ &  $\log\langle
  L_{18}\rangle$ & $\langle N_{18} \rangle$ &
  $\langle N_{19} \rangle$ & $\langle N_{20} \rangle$ \\ 
 $h^{-1}\Msun$ & $h^{-2}\Lsun$ & $h^{-2}\Lsun$ & & & \\
\hline\hline
 $11.56$ &  $9.57$ &  $9.63$ & $0.535$ &   --    &    --   \\ 
 $11.88$ &  $9.77$ &  $9.86$ & $1.238$ & $1.015$ &    --   \\
 $12.20$ & $10.02$ & $10.13$ & $1.416$ & $1.203$ & $0.032$ \\
 $12.50$ & $10.19$ & $10.33$ & $1.879$ & $1.441$ & $0.914$ \\ 
 $12.79$ & $10.32$ & $10.53$ & $2.715$ & $1.904$ & $1.004$ \\ 
 $13.06$ & $10.41$ & $10.69$ & $3.795$ & $2.603$ & $1.385$ \\ 
 $13.31$ & $10.47$ & $10.85$ & $5.311$ & $3.722$ & $1.845$ \\ 
 $13.55$ & $10.53$ & $11.00$ & $8.580$ & $5.258$ & $2.282$ \\ 
 $13.77$ & $10.58$ & $11.15$ & $14.43$ & $7.492$ & $2.947$ \\ 
 $13.96$ & $10.62$ & $11.28$ & $21.55$ & $10.74$ & $3.801$ \\ 
 $14.14$ & $10.67$ & $11.41$ & $28.65$ & $13.95$ & $5.094$ \\ 
 $14.30$ & $10.75$ & $11.52$ & $43.72$ & $19.87$ & $5.903$ \\ 
 $14.44$ & $10.80$ & $11.62$ & $70.00$ & $24.22$ & $8.323$ \\ 
 $14.58$ & $10.85$ & $11.69$ & $73.01$ & $28.08$ & $6.852$ \\ 
 $14.69$ &    --   & $11.77$ & $112.5$ & $25.12$ & $10.08$ \\ 
 $14.80$ &    --   & $11.82$ &   --    & $47.65$ & $7.219$ \\ 
\hline
\end{tabular}
\medskip
\begin{minipage}{\hssize}
  Column (1), (2) and (3)  list the group mass, the average luminosity
  of the  central group galaxy,  and the average, total  luminosity of
  all group  galaxies with $M_{b_J}-5\log  h \leq -18$.   Columns (4),
  (5),  and  (6) list  the  average  number  of galaxies,  per  group,
  brighter   than  $M_{b_J}-5\log   h   =  -18$,   $-19$  and   $-20$,
  respectively.  Note  that the  group masses are  only valid  for the
  WMAP3  cosmology used  here.   However, it  is  straight forward  to
  convert these numbers to any other cosmology.
\end{minipage}
\end{table}

An  obvious shortcoming  of  this  method is  that  the true  relation
between $L_{18}$ and $M$ contains  some scatter, which thus results in
errors in the inferred group masses. However, detailed tests with mock
galaxy  redshift  surveys have  shown  that  this method  nevertheless
allows for a  very accurate recovery of {\it  average} halo occupation
statistics.   In  particular, the  group  finder  yields average  halo
occupation  numbers  and  average  mass-to-light ratios  that  are  in
excellent agreement with the  input values (Yang \etal 2005b; Weinmann
\etal 2006b).  

Application of  this group  finder to the  2dFGRS, yields  a catalogue
consisting  of  $77,708$  groups,  which in  total  contain  $104,912$
galaxies.  Among these, $7251$  are binaries, $2343$ are triplets, and
$2502$ are  systems with four members  or more.  The  vast majority of
the groups  ($66,612$ systems) in  our catalogue, however,  consist of
only a single member.  Note  that some faint galaxies are not assigned
to any group,  because it is difficult to decide  whether they are the
satellite galaxies of larger systems, or the central galaxies of small
haloes.   Table~A1  lists a  number  of  average  properties of  these
groups, as function of the  assigned group mass. These properties have
been used in this paper for comparison with our CLF predictions.

\section[]{Mock Galaxy Redshift Surveys}
\label{sec:AppB}

We construct MGRSs  by populating dark matter haloes  with galaxies of
different  luminosities. The  distribution  of dark  matter haloes  is
obtained from two large  $N$-body simulations of $N=512^3$ dark matter
particles each.  These simulations have  been carried with  PKDGRAV, a
tree-code written  by Joachim Stadel  and Thomas Quinn (Stadel  2001). 
Each simulation  evolves the  distribution of the  dark matter  in the
WMAP3         $\Lambda$CDM         cosmology        ($\Omega_m=0.238$,
$\Omega_{\Lambda}=0.762$, $\Omega_b=0.042$, $h=0.73$, $\sigma_8=0.75$,
$n_s = 0.951$).  The initial conditions are generated with the GRAFIC2
package (Bertschinger 2001), which also computes the transfer function
as described in  Ma \& Bertschinger (1995).  The  two simulations have
periodic boundary  conditions and box  sizes of $100 h^{-1}  \Mpc$ and
$300 h^{-1} \Mpc$, respectively.  The particle masses are $4.92 \times
10^8 \msunh$ and  $1.33\times 10^{10} \msunh$ for the  small and large
box  simulations,  respectively.  In  what  follows  we  refer to  the
simulations with  $L_{\rm box}=100  h^{-1} \Mpc$ and  $L_{\rm box}=300
h^{-1} \Mpc$ as $L_{100}$ and $L_{300}$, respectively.

We follow  Yang \etal (2004) and  replicate the $L_{300}$ box  on a $4
\times 4 \times 4$ grid.  The central $2 \times 2 \times 2$ boxes, are
replaced by a stack of $6  \times 6 \times 6$ $L_{100}$ boxes, and the
virtual observer  is placed at the  center (see Fig.~11  in Yang \etal
2004).  This stacking  geometry circumvents incompleteness problems in
the mock survey  due to insufficient mass resolution  of the $L_{300}$
simulations,  and allows  us to  reach  the desired  depth of  $z_{\rm
  max}=0.20$ in all directions.

Dark  matter haloes are  identified using  the standard  FOF algorithm
with  a  linking  length   of  $0.2$  times  the  mean  inter-particle
separation.  Unbound haloes and haloes with less than 10 particles are
removed  from the  sample. The  resulting halo  mass functions  are in
excellent agreement  with the analytical halo mass  function of Sheth,
Mo \& Tormen (2001). We  populate the dark matter haloes with galaxies
of different luminosity using our CLF.  Because of the mass resolution
of  the simulations  and  because  of the  completeness  limit of  the
2dFGRS, we adopt a minimum  galaxy luminosity of $L_{\rm min} = 10^{7}
h^{-2} \Lsun$. The halo occupation statistics of these galaxies follow
from the CLF as described in \S\ref{sec:Nstat}. Luminosities are drawn
using  either  the `deterministic'  or  the  `random' sampling  method
described  in \S\ref{sec:censat},  whereby we  always assume  that the
central galaxy is the brightest galaxy in its halo.

The positions and velocities of  the galaxies with respect to the halo
center-of-mass are  drawn assuming that  the brightest galaxy  in each
halo resides at  rest at the center.  The  satellite galaxies follow a
number  density distribution  that is  identical to  that of  the dark
matter  particles, and  are  assumed to  be  in isotropic  equilibrium
within the dark matter potential.   To construct MGRSs we use the same
selection criteria  and observational biases as in  the 2dFGRS, making
detailed use of the survey  masks provided by the 2dFGRS team (Colless
\etal 2001; Norberg  \etal 2002).  We also mimic  fiber collisions and
image blending as described in detail  in van den Bosch \etal (2005b). 
The  final  MGRSs  accurately  match the  clustering  properties,  the
apparent magnitude  distribution and the redshift  distribution of the
2dFGRS, and mimic all the various incompleteness effects, allowing for
a direct, one-to-one comparison with the true 2dFGRS.

\label{lastpage}

\end{document}

%% file: macros_vdbosch.tex
\newcommand{\etal}{{et al.~}}

\newcommand{\kmsmpc}{\>{\rm km}\,{\rm s}^{-1}\,{\rm Mpc}^{-1}}
\newcommand{\kms}{\>{\rm km}\,{\rm s}^{-1}}

\newcommand{\Mpc}{\>{\rm Mpc}}
\newcommand{\kpc}{\>{\rm kpc}}
\newcommand{\Msun}{\>{\rm M_{\odot}}}
\newcommand{\Lsun}{\>{\rm L_{\odot}}}
\newcommand{\MLsun}{\>({\rm M}/{\rm L})_{\odot}}

\newcommand{\calM}{{\cal M}}

\newcommand{\walpha}{\tilde{\alpha}}
\newcommand{\wLstar}{\tilde{L}^{*}}

\newcommand{\beq}{\begin{equation}}
\newcommand{\eeq}{\end{equation}}

\newcommand{\rmd}{{\rm d}}

\newcommand{\msunh}{\>h^{-1}\rm M_\odot}

\def\gtsima{$\; \buildrel > \over \sim \;$}
\def\ltsima{$\; \buildrel < \over \sim \;$}
\def\prosima{$\; \buildrel \propto \over \sim \;$}
\def\gsim{\lower.7ex\hbox{\gtsima}}
\def\lsim{\lower.7ex\hbox{\ltsima}}
\def\simgt{\lower.7ex\hbox{\gtsima}}
\def\simlt{\lower.7ex\hbox{\ltsima}}
\def\simpr{\lower.7ex\hbox{\prosima}}
\def\la{\lsim}
\def\ga{\gsim}
\def\lta{\la}
\def\gta{\ga}

\newcommand{\apj}{ApJ}
\newcommand{\apjs}{ApJS}
\newcommand{\aj}{AJ}
\newcommand{\mnras}{MNRAS}
\newcommand{\aap}{A\&A}

\newdimen\hssize
\newdimen\hdsize
